\documentclass[12pt]{article}

\usepackage[dvips]{graphicx}
\usepackage{latexsym}

\usepackage{amssymb,amsfonts,amsmath}
\usepackage{graphicx} 
\usepackage{indentfirst}

 \usepackage{bbm}

\parskip=\medskipamount

\newcommand {\cD}{{\cal D}}

\newcommand {\cF}{{\cal F}}
\newcommand {\cG}{{\cal G}}
\newcommand {\cH}{{\cal H}}

\newcommand {\cL}{{\cal L}}
\newcommand {\cM}{{\cal M}}
\newcommand {\cN}{{\cal N}}
\newcommand {\cO}{{\cal O}}
\newcommand {\cP}{{\cal P}}

\newcommand {\cW}{{\cal W}}

\newcommand{\bE}{{\bf E}}

\newcommand{\bH}{{\bf H}}

\newcommand{\bR}{{\bf R}}

\newcommand{\bT}{{\bf T}}

\def\a{\alpha}
\def \bi{\bibitem}

\def\b{\beta}

\def\d{\delta}
\def\e{\epsilon}
\def\f{\phi}
\def\g{\gamma}
\def\G{\Gamma}

\def\j{\psi}
\def\k{\kappa}
\def\l{\lambda}

\def\o{\omega}

\def\q{\theta}
\def\r{\rho}
\def\s{\sigma}
\def\t{\tau}

\def\x{\xi}
\def\z{\zeta}
\def\D{\Delta}
\def\F{\Phi}
\def\J{\Psi}
\def\L{\Lambda}
\def\O{\Omega}

\def\Q{\Theta}
\def\S{\Sigma}
\def\U{\Upsilon}

\def\rd{{\rm d}}
\def\ri{{\rm i}}
\def\re{{\rm e}}

\newcommand{\ad}{{\dot{\alpha}}}                           
\newcommand{\bd}{{\dot{\beta}}}                            
\newcommand{\ve}{\varepsilon}                            
\newcommand{\cDB}{{\bar\cD}}                            
\newcommand{\DB}{\bar{D}}

\newcommand{\pa}{\partial}                           
\newcommand{\hf}{\frac12}

\newcommand{\vf}{\varphi}
\newcommand{\sect}[1]{\setcounter{equation}{0}\section{#1}}

\newcommand{\be}{\begin{equation}}
\newcommand{\ee}{\end{equation}}
\newcommand{\bea}{\begin{eqnarray}}
\newcommand{\eea}{\end{eqnarray}}
\newcommand{\non}{\nonumber}
\newcommand{\1}{\underline{1}}
\newcommand{\2}{\underline{2}}

\def\dt#1{{\buildrel {\hbox{\LARGE .}} \over {#1}}}    

\newcommand{\bm}[1]{\mbox{\boldmath$#1$}}

\def\double #1{#1{\hbox{\kern-2pt $#1$}}}

\newcommand{\hm}{\hat{m}}
\newcommand{\hn}{\hat{n}}
\newcommand{\ha}{\hat{a}}
\newcommand{\hb}{\hat{b}}
\newcommand{\hc}{\hat{c}}

\newcommand{\hal}{\hat{\a}}
\newcommand{\hbe}{\hat{\b}}
\newcommand{\hga}{\hat{\g}}
\newcommand{\hde}{\hat{\d}}
\newcommand{\hrh}{\hat{\rho}}

\renewcommand{\Bar}{\overline}

\begin{document}

\begin{titlepage}

\begin{flushright}
arXiv:0704.1185 [hep-th] \\
April, 2007\\
\end{flushright}
\vspace{5mm}

\begin{center}
{\Large \bf  Five-dimensional $\bm{\cN=1}$ AdS superspace:}\\ 
{\large\bf Geometry,  off-shell multiplets and dynamics }
\end{center}

\begin{center}

{\large  
Sergei M. Kuzenko\footnote{{kuzenko@cyllene.uwa.edu.au}}
and 
Gabriele Tartaglino-Mazzucchelli\footnote{gtm@cyllene.uwa.edu.au}
} \\
\vspace{5mm}

\footnotesize{
{\it School of Physics M013, The University of Western Australia\\
35 Stirling Highway, Crawley W.A. 6009, Australia}}  
~\\

\vspace{2mm}

\end{center}
\vspace{5mm}

\begin{abstract}
\baselineskip=14pt
As a  step towards formulating projective superspace techniques
for supergravity theories with eight supercharges, this work is devoted to
field theory in five-dimensional $\cN=1$ anti-de Sitter superspace
AdS${}^{5|8}={\rm SU}$(2,2$|$1)/SO$(4,1)\times {\rm U}(1)$ which is  a maximally symmetric
curved background. 
We  develop the differential geometry  of AdS${}^{5|8}$
and describe its isometries in terms of Killing supervectors.
Various off-shell supermultiplets in  AdS${}^{5|8} \times S^2$
are defined, and supersymmetric actions are constructed both in
harmonic and projective superspace approaches.
Several families of supersymmetric theories are presented
including nonlinear sigma-models, 
Chern-Simons theories
 and vector-tensor dynamical systems.
Using a suitable  coset representative, we make use of the coset construction
to develop an explicit realization 
for one half of the superspace AdS${}^{5|8}$
as  a trivial fiber bundle with fibers isomorophic to four-dimensional Minkowski
superspace.
\end{abstract}
\vspace{1cm}

\vfill
\end{titlepage}

\newpage
\renewcommand{\thefootnote}{\arabic{footnote}}
\setcounter{footnote}{0}

\tableofcontents{}
\vspace{1cm}
\bigskip\hrule

\sect{Introduction}

In four-dimensional $\cN=2$ Poincar\'e supersymmetry, 
there exist two powerful formalisms to construct off-shell manifestly supersymmetic actions:
harmonic superspace \cite{HarmonicSuperspace,GIOS} and projective superspace
\cite{ProjectiveSuperspace,SiegelProjective,LR1,LR2}. Both approaches make use of the superspace
${\mathbb R}^{4|8} \times S^2$ and its supersymmetric subspaces, which were 
introduced for the first time by Rosly \cite{Rosly} 
who built on earlier ideas due to Witten \cite{Witten}.
Both approaches can naturally be extended to the case of $d$-dimensional supersymmetry 
with eight supercharges, for $d\leq 6$, where the appropriate flat superspace with 
auxiliary bosonic dimensions is ${\mathbb R}^{d|8} \times S^2$. 
Specifically, the harmonic superspace formulations were developed 
in \cite{Z,KuzLin} for $d=5$, and in \cite{6Dhar} for $d=6$.
The projective superspace formulations were developed 
in \cite{KuzLin} for $d=5$, and in \cite{GL,GPT} for $d=6$.

In projective superspace, off-shell multiplets are reasonably short
and can readily be expressed in terms of 4D $\cN=1$ superfields. 
The latter property is very appealing  from the point of view of brane(-world)  models. 
It is also expected \cite{Siegel,LW} that projective superspace should be relevant 
in the context of hybrid string theory \cite{hybrid}.
${}$For these and similar possible applications, one  actually needs projective superspace 
techniques for supergravity. So far, to the best of our knowledge, the projective superspace 
approach has been mastered only in the flat case.

In harmonic superspace, the prepotential structure of 4D $\cN=2$ supergravity
is well understood \cite{SUGRA-har,GIOS}, and similar constructions 
are clearly applicable in five and six dimensions, see 
\cite{Sokatchev} for the six-dimensional case.
What is still missing here, in our opinion,  is a properly incorporated
covariant formalism of differential geometry 
for superfield supergravity, which should be similar
in  spirit to the famous Wess-Zumino approach 
to (the old minimal formulation for) 
4D $\cN=1$ supergravity reviewed in 
\cite{WessBagger}. In four-dimensional $\cN=1$ supergravity,  
it has been recognized  for a long time that 
the most efficient approach to superfield supergravity 
occurs if one merges together and uses, depending on a concrete application,
 both the  covariant and prepotential techniques 
\cite{GGRS,BK}.

Unlike the purely prepotential approach pursued in \cite{SUGRA-har,GIOS}, 
this paper is targeted at (making the  first step towards) developing 
covariant superfield techniques for supergravity theories with eight supercharges.
Our point of departure is as follows.
It is known that all information about off-shell 
supergravity formulations (including the structure of possible matter 
multiplets) is encoded in the corresponding algebra of covariant derivatives. 
We would like to use only this input and try to develop techniques
to construct supersymmetric actions both in the harmonic and projective settings.
In this paper we consider one particular supergravity background -- 
five-dimensional $\cN=1$ anti-de Sitter superspace, 
AdS${}^{5|8}$, and explicitly develop harmonic and projective formulations 
in a covariant fashion using only the language of differential geometry.
We believe that similar ideas should be applicable 
for a general supergravity background, as well as  in four and six space-time dimensions.
In particular, the case of 4D $\cN=2$ anti-de Sitter superspace\footnote{The 4D
$\cN=2$ anti-de Sitter superspace was studied in detail in \cite{SS} 
where a manifestly supersymmetric formulation 
for the off-shell 4D $\cN=2$ anti-de Sitter higher spin supermultiplets \cite{GKS}
was given. A few years later, some formal aspects of this superspace
were also discussed in \cite{BILS}.}
can be treated similarly.

This paper is organized as follows. In section 2 we derive the algebra of the
covariant derivatives for 5D $\cN=1$ anti-de Sitter superspace 
by solving the Bianchi identities. In section 3 the isometries of AdS${}^{5|8}$ 
are realized in terms of Killing supervectors. In section 4 we introduce 
analytic multiplets over the harmonic superspace AdS${}^{5|8} \times S^2$ 
and formulate the harmonic superspace action. 
Various projective multiplets are defined in section 5, as well as the projective 
superspace action is formulated. A remarkable feature of this supersymmetric action  is that 
it is uniquely determined by two  independent requirements: (i) projective invariance;
(ii) invariance under the isometry group SU$(2,2|1)$. 
Some important examples of dynamical systems in the AdS projective superspace 
are given in section 6. 
An explicit  coset construction 
for one half of AdS${}^{5|8}$ (Poincar\'e chart) is elaborated in section 7.
Our 5D notation and conventions are collected in Appendix A.

\sect{Covariant derivatives}
\label{sectCovariantDerivatives}

In this section, we develop the differential geometry  
of five-dimensional $\cN=1$ anti-de Sitter superspace, 
AdS${}^{5|8}$. This is a supersymmetric version of spaces of constant curvature and, 
similar to all symmetric spaces, it can be realized as a coset space, specifically
AdS${}^{5|8}={\rm SU}$(2,2$|$1)/SO(4,1)$\times$U(1). Group-theoretical aspects
of AdS${}^{5|8}$ will be discussed in section 7.

Let $z^{\hat{M}}=(x^{\hm},\q^{\hat{\mu}}_i)$
be local bosonic ($x$) and fermionic ($\q$) 
coordinates parametrizing  AdS${}^{5|8}$,
where $\hm=0,1,\cdots,4$, $\hat{\mu}=1,\cdots,4$, and  $i=\1,\2$.
The Grassmann variables $\q^{\hat{\mu}}_i$
are assumed to obey
a standard pseudo-Majorana reality condition. 
Since the holonomy group of AdS${}^{5|8}$ is ${\rm SO}(4,1)\times {\rm U}(1)$,
the superspace covariant derivative 
$\cD_{\hat{A}} =(\cD_{\hat{a}}, \cD_{\hat{\a}}^i)$
can be chosen to have the form 
\be
\cD_{\hat{A}}~=~E_{\hat{A}}~+~{\rm i}\,\Phi_{\hat{A}}\,J~+~\hf \,\O_{\hat{A}}{}^{\hb\hc}\,M_{\hb\hc}
~=~E_{\hat{A}}~+~{\rm i}\,\Phi_{\hat{A}}\,J~+~\O_{\hat{A}}{}^{\hbe\hga}\,M_{\hbe\hga}~.\label{CovDev}
\ee
Here $E_{\hat{A}}= E_{\hat{A}}{}^{\hat{M}}(z) \pa_{\hat{M}}$ is the supervielbein, 
with $\pa_{\hat{M}}= \pa/ \pa z^{\hat{M}}$,
$J$ the Hermitian  generator of the group U(1), 
$M_{\hb\hc}$ the generators of the Lorentz group ${\rm SO}(4,1)$, 
and $\Phi_{\hat{A}}(z)$  and $\O_{\hat{A}}{}^{\hb\hc}(z)$ the corresponding connections. 
The Lorentz generators with vector indices ($M_{\ha\hb}=-M_{\hb\ha}$) and spinor indices
($M_{\hal\hbe}=M_{\hbe\hal}$) are related to each other by the rule:
$M_{\ha\hb}=(\S_{\ha\hb})^{\hal\hbe}M_{\hal\hbe}$, see Appendix A for more details
regarding our 5D notation and conventions.
The generators of the holonomy group
act on the covariant derivatives as follows:
\bea
{[}J,\cD_{\hal}^i{]}&=&J^i_{~j}\cD_{\hal}^j~,\\
{[}M_{\hal\hbe},\cD_{\hga}^i{]}&=&{1\over 2}\Big(\ve_{\hga\hal}\cD_{\hbe}^i 
+ \ve_{\hga\hbe}\cD_{\hal}^i\Big)~.
\eea
The Hermitian matrix $J^i_{~j}$ 
should be 
traceless, $J^{i}{}_{i}=0$, in order to preserve the
pseudo-Majorana condition enjoyed by the covariant derivatives. 
The latter condition is equivalent to the fact that 
the isotensors\footnote{Two-component indices  $i,~j$ are raised and lowered 
using the SL$(2,{\mathbb C})$-invariant antisymmetric tensors $\ve^{ij}$ and $\ve_{ij}$
normalized by $\ve^{ik} \ve_{kj}=\d^i_j$ and $\ve^{\1\2}=\ve_{\2\1}=1$. }
$J^{ij}=\ve^{jk}J^i_{~k}$ and $J_{ij}=\ve_{ik}J^k_{~j}$ 
are symmetric,
$J^{ij}=J^{ji}$, $J_{ij}=J_{ji}$.
The fact that  $J^i_{~j}$ is Hermitian, can be seen to be  equivalent to
 $(J^{ij})^*=-J_{ij}$.

The algebra of covariant derivatives can be reconstructed if we impose
the following two requirements:
(i) the torsion tensor is covariantly constant;\footnote{Then, in accordance 
with Dragon's theorem \cite{Dragon}, the curvature tensor is covariantly constant.}  
(ii) the group ${\rm SO}(4,1)\times {\rm U}(1)$ belongs to the automorphism  group.
These requirements lead, in particular, to the ans\"atze:
\bea
\{\cD_{\hal}^i,\cD_{\hbe}^j\}&=&-2{\rm i}\ve^{ij}\cD_{\hal\hbe}
+x\ve^{ij}\ve_{\hal\hbe}J+f^{ij}M_{\hal\hbe}~,
\label{ansatz1}\\
{[} \cD_{\ha}, \cD_{\hal}^i {]}
&=&C^i_{~j}(\G_{\ha})_{\hal}^{~\hbe}\cD^j_{\hbe}~,
\label{ansatz2}
\eea
where 
\be
\cD_{\hal\hbe}=(\G^{\ha})_{\hal\hbe}\cD_{\ha}~,
\ee
and $x$ is a constant parameters, $f^{ij}=f^{ji}$, 
$C^i_{~j}$ is a  $2\times 2$ matrix. 
Eq. (\ref{ansatz2}) can be  rewritten in the equivalent form
\bea
{[} \cD_{\hal\hbe}, \cD_{\hga}^i {]}
&=&-2C^i_{~j}\Big(\ve_{\hga\hal}\cD^j_{\hbe}-\ve_{\hga\hbe}\cD^j_{\hal}
+{1\over2}\ve_{\hal\hbe}\cD^j_{\hga}\Big)~,
\label{ansatz2.2}
\eea
Note that setting $x=m=C^i{}_{j}=0$  gives 
the flat supersymmetry algebra, see. e.g.,  \cite{KuzLin}.

The (covariantly) constant parameters $x$, $f^{ij}$ and $C^i_{~j}$
in (\ref{ansatz1}) and (\ref{ansatz2}) 
turn out  to be considerably constrained on general grounds.
Firstly, the tensor
$f^{ij}$ must be invariant under the action of $J$,
\bea
Jf^{ij}\,=\,(J^i_{~k}f^{kj}+J^j_{~k}f^{ik})\,=\,0
\quad
\Longleftrightarrow \quad 
f^{ij}~=~m J^{ij}~,~~~~~
\eea
with $m$ a constant parameter.
Secondly, we should take care of reality conditions such as
\be
(\cD_{\hal}^i F)^*=-(-1)^{\e(F)}\cD^{\hal}_i F^*~,
\ee
where $\e(F)$ is the Grassmann parity of $F$. 
They imply that
\bea
x~=~\Bar{x}~, \qquad m~=~\Bar{m}~.
\eea
and $C^i{}_j$ is anti-Hermitian,
\be
C^\dagger = -C~, \qquad C =( C^i{}_j)~.
\ee

Of course,  we should also guarantee the fulfillment of 
the Bianchi identities, and this proves to  lead to additional restrictions
on the parameters.
In particular, the dimension-$3/2$ Bianchi identity
\be
{[}\cD_{\hal}^i,\{\cD_{\hbe}^j,\cD_{\hga}^k\} {]}+{[}\cD_{\hbe}^j,\{\cD_{\hga}^k,\cD_{\hal}^i\} {]}
+{[}\cD_{\hga}^k,\{\cD_{\hal}^i,\cD_{\hbe}^j\} {]}~=~0~
\label{eqBI1}
\ee
can be shown to imply
\bea
C^i_{~j}={{\rm i}\over 2}\o J^i_{~j}~, \qquad \o=\Big({m\over 2}-x\Big)~.
\label{eqomega}
\eea
Imposing the dimension-2 Bianchi identity
\be
{[}\cD_{\ha},\{\cD_{\hal}^i,\cD_{\hbe}^j\} {]}+\{\cD_{\hal}^i,{[}\cD_{\hbe}^j,\cD_{\ha}{]} \}
-\{\cD_{\hbe}^j,{[}\cD_{\ha},\cD_{\hal}^i{]} \}
~=~0~
\label{eqBI2}
\ee
leads, in particular,  to 
\bea
{[}\cD_{\ha},\cD_{\hb}{]}&=&-{\ri\over 16}\ve_{ij}
(\G_{\hb})^{\hal\hbe}{[}\cD_{\ha},\{\cD_{\hal}^i,\cD_{\hbe}^j\}{]}\non\\
&=&{\ri\over 16}\ve_{ij}(\G_{\hb})^{\hal\hbe}
\Big(\{\cD_{\hal}^i,{[}\cD_{\hbe}^j,\cD_{\ha}{]} \}-\{\cD_{\hbe}^j,{[}\cD_{\ha},\cD_{\hal}^i{]}\}\Big)~,
\eea
and then
\bea
{[}\cD_{\ha},\cD_{\hb}{]}&=&{1\over 4}m\o J^2M_{\ha\hb}~,
\label{eqBI2.2}
\eea
where 
\be
J^2~\equiv~-\hf J^{ij}J_{ij}~.
\ee
Another consequence of the dimension-2 Bianchi identity (\ref{eqBI2}) is
\be
\o~=~-{1\over 4}\,m~.
\ee
As a result, all the parameters in (\ref{ansatz1}) and (\ref{ansatz2}) 
have been expressed in terms of $\o$.
With the above conditions taken into account, the remaining
dimension-${5\over 2}$ Bianchi identity
\be
{[}\cD_{\ha},{[}\cD_{\hb},\cD_{\hal}^i{]} {]}+{[}\cD_{\hb},{[}\cD_{\hal}^i,\cD_{\ha}{]} {]}
+{[}\cD_{\hal}^i,{[}\cD_{\ha},\cD_{\hb}{]} {]}
~=~0~,
\label{eqBI3}
\ee
and dimension-3 Bianchi identity
\be
{[}\cD_{\ha},{[}\cD_{\hb},\cD_{\hc}{]} {]}+{[}\cD_{\hb},{[}\cD_{\hc},\cD_{\ha}{]} {]}
+{[}\cD_{\hc},{[}\cD_{\ha},\cD_{\hb}{]} {]}
~=~0~,
\label{eqBI4}
\ee
are satisfied identically.

Let us summarise the results obtained.
The covariant derivatives for   AdS${}^{5|8}$ obey the algebra
\begin{subequations}
\bea
\{\cD_{\hal}^i,\cD_{\hbe}^j\}&=&-2{\rm i}\ve^{ij}\cD_{\hal\hbe}
-3\o\ve^{ij}\ve_{\hal\hbe}J-4\o J^{ij}M_{\hal\hbe}~,
\label{algebra1}\\
{[} \cD_{\ha}, \cD_{\hbe}^i {]}
&=&{\ri\over 2}\o J^i_{~j}(\G_{\ha})_{\hbe}^{~\hga}\cD^j_{\hga}~,
\label{algebra2}\\
{[} \cD_{\ha},\cD_{\hb} {]}&=&-\o^2 J^2 M_{\ha\hb}~.
\label{algebra3}
\eea
\end{subequations}
It is  useful to rewrite (\ref{algebra2}) in the equivalent form
\bea
{[}\cD_{\hal\hbe},\cD^i_{\hga}{]}&=&-\ri\, \o J^i_{~j}\left(\ve_{\hga\hal}\cD^j_{\hbe}
-\ve_{\hga\hbe}\cD^j_{\hal}+{1\over 2}\ve_{\hal\hbe}\cD^j_{\hga}\right)~.
\label{algebra2.2}
\eea
As follows from (\ref{algebra3}), the bosonic body of the  superspace 
is characterised by a  constant negative curvature, and therefore it  is AdS$_5$. 
Indeed, since  $J^i_{~j}$ is Hermitian and traceless, we have
\bea
J^i{}_{j}~=~J^I(\s_I)^i_{~j}~&\Longrightarrow&~
J^2=-{1\over 2}J^{ij}J_{ij}
=\hf J^IJ^J\,{\rm tr}(\s_I\s_J)=J^I J_I~>~0
~,~~~
\eea 
where  $J^I$ is a real tree-vector, with $I=1,\,2,\,3$, and $\s^I$ are the Pauli matrices. 
In section \ref{CosetSection},  we will give  an explicit  
(coset space) realization of the geometry described.

Up to an isomorphism, one can always choose 
$J^i{}_j \propto (\s_3 )^i{}_j$, and hence $J^{\1 \1}=J^{\2 \2} =0$.
Then, it follows from (\ref{algebra1}--\ref{algebra3}) that each 
of the two subsets of covariant derivatives $(\cD_{\hat a},  \cD^{\1}_{\hat \a})$
and  $(\cD_{\hat a},  \cD^{\2}_{\hat \a})$ forms a closed algebra, in particular
\begin{subequations}
\bea
\{\cD_{\hal}^{\1},\cD_{\hbe}^{\1}\}&=&0~,
\label{algebrareduced1}\\
{[} \cD_{\ha}, \cD_{\hbe}^{\1} {]}
&=&{\ri\over 2}\o J^{\1}{}_{\1}(\G_{\ha})_{\hbe}{}^{\hga}\cD^{\1}_{\hga}~,
\label{algebrareduced2}\\
{[} \cD_{\ha},\cD_{\hb} {]}&=&-\o^2 J^2 M_{\ha\hb}~.
\label{algebrareduced3}
\eea
\end{subequations}
Therefore, one can consistently define covariantly chiral 
superfields obeying the constraint $\cD^{\2}_{\hat \a} \F =0$.
Unlike the case of 4D $\cN=1$ anti-de Sitter superspace \cite{IS},
such multiplets can transform in arbitrary representations of 
the  Lorentz group.

In what follows, it will be useful to deal with a different basis
for the spinor covariant derivatives.
Let us introduce 
two linearly independent isospinors $u^+_i$ and $u^-_i$, 
\be
u^{+i}u^-_i~\equiv~(u^+u^-)~\ne~0~~~\Longrightarrow~~~\d^i_j
~=~{1\over (u^+u^-)}(u^{+i}u^-_j-u^{-i}u^+_j)~,
\label{isospinor1}
\ee
which do not transform under the action of $J$, that is
$J \,u^+_i = J\,u^-_i =0$.
Then,  defining
\bea
\cD^{\pm}_{\hal}&\equiv&\cD^i_{\hal}u^{\pm}_i~,\\
J^{++}~\equiv~J^{ij}u^+_iu^+_j~, \qquad J^{+-}&\equiv&J^{ij}u^+_iu^-_j~,
\qquad J^{--}~\equiv~J^{ij}u^-_iu^-_j~.
\eea
the relations  (\ref{algebra1}) and  (\ref{algebra2}) become
\begin{subequations}
\bea
\{\cD_{\hal}^+,\cD_{\hbe}^+\}&=&-4\o J^{++}M_{\hal\hbe}~,
\label{algebra1++}\\
\{\cD_{\hal}^+,\cD_{\hbe}^-\}&=&2(u^+u^-){\rm i}\cD_{\hal\hbe}+3(u^+u^-)\o\ve_{\hal\hbe}J
-4\o J^{+-}M_{\hal\hbe}~~~,\label{algebra1+-}\\
\{\cD_{\hal}^-,\cD_{\hbe}^-\}&=&-4\o J^{--}M_{\hal\hbe}~~,
\label{algebra1--}\\
{[} \cD_{\ha}, \cD_{\hal}^+ {]}
&=&-{\ri \o\over 2(u^+u^-)}(\G_{\ha})_{\hal}^{~\hbe}(J^{++}\cD^-_{\hbe}-J^{+-}\cD^+_{\hbe})~,
\label{algebra2+}\\
{[} \cD_{\ha}, \cD_{\hal}^-{]}
&=&{\ri \o  \over 2(u^+u^-)}(\G_{\ha})_{\hal}^{~\hbe}(J^{--}\cD^+_{\hbe}-J^{+-}\cD^-_{\hbe})~.
\label{algebra2-}
\eea
\end{subequations}
Eqs. (\ref{algebra2+}) and (\ref{algebra2-}) are equivalent to
\begin{subequations}
\bea
{[}\cD_{\hal\hbe},\cD^+_{\hga}{]}&=&{\ri\o\over(u^+u^-)} 
J^{++}\left(\ve_{\hga\hal}\cD^-_{\hbe}-\ve_{\hga\hbe}\cD^-_{\hal}
+{1\over 2}\ve_{\hal\hbe}\cD^-_{\hga}\right)\non\\
&&-{\ri\o\over(u^+u^-)} J^{+-}\left(\ve_{\hga\hal}\cD^+_{\hbe}-\ve_{\hga\hbe}\cD^+_{\hal}
+{1\over 2}\ve_{\hal\hbe}\cD^+_{\hga}\right)~,
\label{algebra2+.2}\\
{[}\cD_{\hal\hbe},\cD^-_{\hga}{]}&=&{\ri\o\over(u^+u^-)} J^{+-}\left(\ve_{\hga\hal}\cD^-_{\hbe}
-\ve_{\hga\hbe}\cD^-_{\hal}+{1\over 2}\ve_{\hal\hbe}\cD^-_{\hga}\right)\non\\
&&-{\ri\o\over(u^+u^-)} J^{--}\left(\ve_{\hga\hal}\cD^+_{\hbe}-\ve_{\hga\hbe}\cD^+_{\hal}
+{1\over 2}\ve_{\hal\hbe}\cD^+_{\hga}\right)~.
\label{algebra2-.2}
\eea
\end{subequations}

Under general coordinate and local SO(4,1)$\times$U(1) transfomations, 
the covariant derivatives change as 
\be
\cD_{\hat{A}} \to \cD'_{\hat{A}}  ={\rm e}^{\t  }\, \cD_{\hat{A}}\, {\rm e}^{-\t  } ~,
\qquad \t = \t^{\hat{B}}(z) \cD_{\hat{B}} 
+{\rm i}\,\t(z) J+\t^{\hbe\hga}(z) M_{\hbe\hga}~.
\ee
This gauge freedom can be used to impose a suitable Wess-Zumino gauge. 
The latter can be chosen such that 
\be
\cD_{\hat{a}} | = \nabla_{\hat a} = e_{\hat a}{}^{\hat m} (x) \, \pa_{\hat m} 
+ \hf \o_{\hat a}{}^{\hat b \hat c} (x) \,M_{\hat  b \hat c}~,
\label{WZ}
\ee
where $U|$ means the $\q$ independent part of a superfield $U(x,\q)$, 
\be
U=U(z) = U(x,\q)~,\qquad U|= U(x,\q=0)~.
\label{projection}
\ee
and  $\nabla_{\hat a}$ stands for  the covariant derivatives
of anti-de Sitter space, 
\bea
{[} \nabla_{\ha},\nabla_{\hb} {]}&=&-\o^2 J^2 M_{\ha\hb}~.
\eea

\sect{Killing supervectors}
\label{sectKillingSupervectors}

Similar to the 4D $\cN=1$ case \cite{BK},
the isometry group ${\rm SU}(2,2|1)$
of ${\rm AdS}^{5|8}$ is generated by those supervector fields
$ \x^{\hat{A}}(z) E_{\hat{A}}$ which enjoy the property 
\bea
\d_\x\cD_{\hat{A}}&=&-{[}(\x
+{\rm i}\r J+\L^{\hbe\hga}M_{\hbe\hga}),\cD_{\hat{A}}{]} =0~,
\label{killing2}
\eea
where
\be
\x \equiv\x^{\hat{A}}\cD_{\hat{A}}=\x^{\ha}\cD_{\ha}+\x^{\hal}_i\cD^i_{\hal}=
-{1\over4}\x^{\hal\hbe}\cD_{\hal\hbe}+\x^{\hal}_i\cD^i_{\hal}~,
\label{killing1}
\ee
for some real scalar $\r(z)$ and symmetric tensor 
$\L^{\hbe\hga}(z)=\L^{\hga\hbe}(z)$.
The $ \x^{\hat{A}}(z) E_{\hat{A}}$
is called a Killing supervector. The set of all Killing supervectors 
forms a Lie algebra with respect to the Lie bracket. Given a Killing 
supervector, 
it generates a symmetry transformation of matter superfields, 
which live on ${\rm AdS}^{5|8}$, defined as
\bea
\d_\x\chi&=&-(\x
+{\rm i} \,\r J+\L^{\hal\hbe}M_{\hal\hbe})\chi~.
\label{killing3}
\eea

Using the (anti) commutation relations (\ref{algebra1}) -- (\ref{algebra3}),
eq. (\ref{killing2}) 
can be seen to be equivalent to
\bea
0~&=&~\Big(\,{1\over 4}\cD_{\hal}^i\x^{\hbe\hga}
+2{\ri}\x^{i\hbe}\d_{\hal}^{\hga}\Big)\cD_{\hbe\hga}
+\,\Big({\ri\over 2}\o\,\x_{\hal}^{~\hbe}J^i_{~j}-\cD^i_{\hal}\x^{\hbe}_j
+{\rm i}\,\r J^i_{~j}\d^{\hbe}_{\hal}+\d^i_j\L_{\hal}^{~\hbe}\Big)\cD_{\hbe}^j~~~\non\\
&&-\,\Big(\,3\o\,\x_{\hal}^i+{\rm i}\cD_{\hal}^i\r\Big)J\,
-\,\Big(\,2\o J^{ij}(\x_j^{\hbe}\d^{\hga}_{\hal}
+\x_j^{\hga}\d^{\hbe}_{\hal})+\cD_{\hal}^i\L^{\hbe\hga}\Big)M_{\hbe\hga}~,
\eea
and from here we deduce 
the set of Killing supervector equations
\begin{subequations}
\bea
\cD_{\hal}^i\x^{\ha}&=&-2{\ri}(\G^{\ha})_{\hal\hbe}\,\x^{i\hbe}~,
\label{EqKilling1}\\
0~&=&{\ri\over 2}\o\,\x_{\hal\hbe}J^i_{~j}-\cD^i_{\hal}\x_{j\hbe}
-{\rm i}\,\r J^i_{~j}\ve_{\hal\hbe}+\d^i_j\L_{\hal\hbe}~,
\label{EqKilling2}\\\
{\rm i} \,\cD_{\hal}^i\r&=&-3\o\,\x_{\hal}^i~,
\label{EqKilling3}\\
\cD_{\hal}^i\L^{\hbe\hga}&=&-2\o J^{ij}(\x_j^{\hbe}\d^{\hga}_{\hal}
+\x_j^{\hga}\d^{\hbe}_{\hal})~.
\label{EqKilling4}
\eea
\end{subequations}
Note that 
(\ref{EqKilling2}) is equivalent to the following equations
\begin{subequations}
\bea
\cD^i_{\hal}\x_{i\hbe}&=&2\L_{\hal\hbe}~,\label{EqKilling2.1}\\
\cD^{i\hal}\x^j_{\hal}+\cD^{j\hal}\x^i_{\hal}&=&8{\rm i}\,J^{ij}\r~,
\label{EqKilling2.2}\\
(\G_{\ha})^{\hal\hbe}(\cD^i_{\hal}\x^j_{\hbe}+\cD^j_{\hal}\x^i_{\hbe})
&=&-4\ri\o\,J^{ij}\x_{\ha}~.
\label{EqKilling2.3}
\eea
\end{subequations}
It is seen that the parameters of U(1) and Lorentz transformations, 
$\r$ and $\L_{\hal\hbe}$, are uniquely expressed in terms of the spinor 
components of the Killing supervector.
As to the vector components $\x^{\ha}$ of $\x$,which is also 
uniquely determined in terms of  the spinor 
components of $\x$, it obeys 
the standard Killing equation 
\be
\cD^{(\ha}\x^{\hb)}\,=\,0~.
\label{EqKilling0}
\ee 
To prove (\ref{EqKilling0}), it suffices to represent $\cD^{\ha}$ in 
(\ref{EqKilling0}) in the form
\be
\cD^{\ha}\,=\,{\ri\over 8}(\G^{\ha})^{\hal\hbe}\cD^i_{\hal}\cD_{i\hbe}~,
\ee
and then make use of relations (\ref{EqKilling1}) and (\ref{EqKilling2.1}).

As is seen from eqs. (\ref{EqKilling3}), (\ref{EqKilling2.1})
 and (\ref{EqKilling2.3}),
the components of $\x$ (hence,  the Lorentz  parameter
$\L_{\hal\hbe}$ as well)
can be expressed in terms of the scalar parameter $\r$ as follows:
\begin{subequations}
\bea
\x_{\hal}^i&=&-{\ri \over 3\o}\,\cD_{\hal}^i\r~,\label{EqRho1}\\
\x_{\ha}&=&-{\ri\over \o J^2}J_{ij}(\G_{\ha})^{\hal\hbe}\,\cD^i_{\hal}\x^j_{\hbe}\,=\,
-{1 \over 3\o^2J^2}J_{ij}(\G_{\ha})^{\hal\hbe}\,\cD^i_{\hal}\cD^j_{\hbe}\,\r
\label{EqRho3}~\\
\L_{\hal\hbe}&=&{1\over 2}\,\cD^i_{\hal}\x_{i\hbe}\,=\,
-{\ri \over 6\o}\,\cD^i_{\hal}\cD_{i\hbe}\,\r~.
\label{EqRho2}
\eea
\end{subequations}
This is similar to the situation in 4D $\cN=2$ AdS supersymmetry \cite{GKS}.

We should point out  that equation (\ref{EqRho2}) implies
\be
0\,=\,\cD^{i\hal}\x_{i\hal}\,=\cD^{i\hal}\cD_{i\hal}\,\r~.
\ee
${}$Furthermore, equations (\ref{EqKilling1}), (\ref{EqKilling4}) 
and (\ref{EqKilling2.2}) imply
\begin{subequations}
\bea
0&=&{1\over \o J^2}J_{jk}(\G^{\ha})^{\hbe\hga}\cD^{i}_{\hal}\cD^j_{\hbe}\x^k_{\hga}
-2(\G^{\ha})_{\hal\hbe}\x^{i\hbe}
~,\label{EqRho4}\\
0&=&{1\over 2}\cD_{\hal}^i\cD^{j\hbe}\x_j^{\hga}
+2\o J^{ij}(\x_j^{\hbe}\d^{\hga}_{\hal}+\x_j^{\hga}\d^{\hbe}_{\hal})~,
\label{EqRho5}\\
0&=&{1\over 3\o}(\cD^{i\hal}\cD^j_{\hal}+\cD^{j\hal}\cD^i_{\hal})\r+8J^{ij}\r~.
\label{EqRho6}
\eea
\end{subequations}
${}$From (\ref{EqRho5}) we also deduce
\bea
\cD_{\hal}^i\cD^{j\hbe}\x_j^{\hga}-\cD_{\hal}^i\cD^{j\hga}\x_j^{\hbe}\,=\,0
\quad \Longrightarrow \quad
\cD_{\hal}^i\cD^{\hbe\hga}\r\,=\,0~, 
\eea
and hence
\be
\cD_{\hbe\hga}\,\r\,=\,0~.
\label{covConst}
\ee
We conclude that $\r$ is
annihilated by the vector covariant derivatives.

${}$For later applications, we 
also observe that 
the relation 
$\cD_{\hal}^i\x^{\hal}_i=0$ 
and eq.  (\ref{EqKilling1}) imply
\be
\cD_{\hal\hbe}\,\x^{i\hbe}\,=\,{5\over 2}\,\ri\,\o J^i_{~j}\x^j_{\hal}~.
\label{KillingDiracEq}
\ee

\sect{Harmonic superspace approach}
\label{sectHarmonicSuperspace}

In the previous two sections, we have described the differential geometry 
of five-dimensional  $\cN=1$ AdS superspace and its isometries.
${}$From now on, we turn to constructing off-shell supersymmetric theories in AdS${}^{5|8}$.
This section is devoted to developing a harmonic superspace approach.
To comply with the conventions 
generally accepted by
the harmonic superspace practitioners
\cite{HarmonicSuperspace,GIOS}, the isospinors $u^+$ and $u^-$ in 
(\ref{algebra1++}--\ref{algebra2-}) 
will be chosen to obey the following constraints:
\bea
(u_i{}^-\,,\,u_i{}^+)\in {\rm SU(2)}~, \qquad (u^{+i})^*=u^-_i~, \qquad
(u^+u^-)=1~.
\label{SU(2)}
\eea
As a first step, 
it is natural to introduce analytic supermultiplets living on harmonic superspace.

\subsection{Analytic multiplets} 

We start our analysis with the introduction of $O(n)$ supermultiplets living in  AdS${}^{5|8}$.
Such a multiplet is described by 
a completely symmetric  superfield $H^{i_1\cdots i_n}(z)=H^{(i_1\cdots i_n)}(z)$ 
(with the symmetrization involving  a factor of ${1/n!}$) 
constrained to enjoy the analyticity condition\footnote{In 4D $\cN=2$ supersymmetry, 
off-shell superfields  $H^{(i_1\cdots i_n)}(z)$ obeying the constrains
$D_{ \a}{}^{(i_1}H^{i_2\cdots i_{n+1})}(z)
= {\bar D}_{ \ad}{}^{(i_1}H^{i_2\cdots i_{n+1})}(z)=0$
have a long history. 
In the presence of an intrinsic central charge, 
the cases $n=1$ and $n=2$ correspond to the Fayet-Sohnius hypermultiplet 
\cite{Sohnius} and the linear multiplet \cite{BS} respectively. 
In the absence of central charge, the case $n=2$ corresponds to the tensor multiplet
\cite{N=2tensor}. The case $n=4$ was discussed in \cite{SSW}.
The multiplets with $n>2$ were introduced in \cite{ProjectiveSuperspace2}, 
in the projective superspace approach, and then re-discovered in  
\cite{LR1}. They were called ``$O(n)$ multiplets'' in \cite{G-RLRvUW}.
Their harmonic superspace description was given in \cite{GIO}.
}
\be
\cD_{\hat \a}{}^{(i_1}H^{i_2\cdots i_{n+1})}(z)\,=\,0~.
\label{O(n)1}
\ee
It follows from the algebra of covariant derivatives, 
that this constraint is consistent provided
the superfield is scalar with respect to SO(4,1). 
If one associates with $H^{i_1\cdots i_n}(z)$ 
a superfield $H^{(n)}(z,u)$ of harmonic charge $n$, 
\be
H^{(n)}(z,u)\,=\,u^+_{i_1}\cdots u^+_{i_n}\,H^{i_1\cdots i_n}(z)~,
\label{O(n)2}
\ee
the analyticity condition (\ref{O(n)1}) 
can be seen to be equivalent to
\be
\cD^+_{\hal}H^{(n)}(z,u)\,=\,0~, \qquad D^{++}H^{(n)}(z,u)\,=\,0~.
\label{O(n)2con}
\ee
Here $D^{++}$ is one of the harmonic derivatives 
$(D^{++}, D^{--}, D^0)$,
\bea
D^{++}=u^{+i}\frac{\partial}{\partial u^{- i}} ~,\quad
D^{--}=u^{- i}\frac{\partial}{\partial u^{+ i}} ~,\quad
D^0&=&u^{+i}\frac{\partial}{\partial u^{+i}}-u^{-i}
\frac{\partial}{\partial u^{-i}} ~,\non \\
{[}D^0,D^{\pm\pm}]=\pm 2D^{\pm\pm}~, \qquad  [D^{++},D^{--}]&=&D^0~,
\label{5}
\eea
which form a basis in the space of left-invariant vector fields for SU(2).

Without imposing the analyticity condition, eq.  (\ref{O(n)1}),
one can consistently define an isotensor  
superfield $F^{i_1\cdots i_n}(z)=F^{(i_1\cdots i_n)}(z)$
that transforms under the action of the isometry group as follows: 
\be
\d_{\x}F^{i_1\cdots i_n}\,=\,
-\Big(\x+{\rm i}\, \r J\Big) \,F^{i_1\cdots i_n}
\,=\,-\x F^{i_1\cdots i_n}
- {\rm i}\,q \,n \, \r J^{(i_1}_{~~\,k}F^{i_2\cdots i_{n})k}~,
\ee
where 
$\x$ is the Killing supervector, and $q$ is 
the $J$-charge of $F^{i_1\cdots i_n}$. 
One  can associate with $F^{i_1\cdots i_n}(z) $ the harmonic superfield
$F^{(n)}(z,u)\,=\,u^+_{i_1}\cdots u^+_{i_n}F^{i_1\cdots i_n}(z)$. 
The latter obeys the algebraic constraint $D^{++}F^{(n)}=0$, and its  
isometry transformation is
\bea
\d_{\x}F^{(n)}&=&-
\x
F^{(n)}
+{\rm i}\, q \r ( 
 J^{++}D^{--}F^{(n)}-
n J^{+-}F^{(n)} )
~,
\eea
where it has been used the fact that $u^\pm_i$ 
are inert under the action of $J$. It is also worth noting that the Killing supervector 
can be rewritten as 
\be
\x= \x^{\ha}\cD_{\ha} 
-\x^{+\hal}\cD^-_{\hal} +\x^{-\hal}\cD^+_{\hal}~.
\ee
It is easy to see that the 
constraint $D^{++}F^{(n)}=0$ is preserved under the isometry 
transformations $D^{++}\d_\x F^{(n)}=0$.

If the superfield  $F^{(n)}$ is constrained to be analytic, $\cD^+_{\hal}F^{(n)}=0$, 
then the value of its $J$-charge turns out to be uniquely fixed, and namely   $q=1$. 
Therefore, the isometry transformation of the $O(n)$ multiplet is 
\bea
\d_{\x}H^{(n)}&=&-\Big(\x+{\rm i}\, \r J\Big) \,H^{(n)}
=-\Big(\x^{\ha}\cD_{\ha}-\x^{+\hal}\cD^-_{\hal}
-{\rm i}\,\r \big( J^{++}D^{--} -n J^{+-} \big)\Big)H^{(n)}~.~~~~~~~~~
\eea

It is not difficult to extend 
the above consideration 
to include more general 
multiplets. Within the harmonic superspace approach \cite{GIOS},
one has to deal with superfields of the form $Q^{(n)}(z,u)$, with $n$ an integer, 
 such that 
(i) $Q^{(n)}(z,u)$ is a smooth function over the group manifold SU(2) 
parametrized by $u=(u_i{}^-,u_i{}^+)$; (ii) under harmonic phase transformations
$u^\pm \to \exp (\pm {\rm i} \vf ) u^\pm$, 
the charge of $Q^{(n)}(z,u)$ is equal to $n$,
$$Q^{(n)}(z,{\rm e}^{ {\rm i} \vf } u^+, {\rm e}^{ -{\rm i} \vf } u^-)
={\rm e}^{n {\rm i} \vf }\, 
Q^{(n)}(z,u^+,  u^-)~, \quad \Longleftrightarrow \quad
D^0 Q^{(n)}(z,u)= nQ^{(n)}(z,u)~.
$$  
Such a superfield can be represented by a convergent Fourier series
(for definiteness, we choose $n\geq 0$)
\bea
Q^{(n)}(z,u)&=&\sum_{k=0}^{+\infty}Q^{(i_1\cdots i_{k+n}j_1\cdots j_k)}
(z)\,u^+_{i_1}\cdots u^+_{i_{k+n}}u^-_{j_1}\cdots u^-_{j_k}~.
\label{HarmonicSeries}
\eea
To realise an action of the U(1) generator $J$ on  $Q^{(n)}$, we define
the component superfields in (\ref{HarmonicSeries})
to transform by the law: 
\be
JQ_{q}^{(i_1\cdots i_{k+n}j_1\cdots j_k)}
=q(2k+n) J^{(i_1}_{~~\,r}Q_{q}^{i_2\cdots i_{k+n}j_1\cdots j_k)r}~, 
\ee 
with the same charge $q$ for all the component superfields.
This leads to
\bea
J \,Q^{(n)}_{q}(z,u)&=&q\Big(J^{--}D^{++}
-J^{++}D^{--}+nJ^{+-}\Big) Q^{(n)}_{q}(z,u)~.~~~~
\eea
The $J$-charge turns out to be uniquely fixed, $q=1$,
if $Q^{(n)}_{q}$ is covariantly analytic, $\cD^+_{\hal} Q_{q}^{(n)}=0$.

To summarise, given a covariantly analytic superfield $Q^{(n)}(z,u)$,
\bea
\cD^+_{\hal} Q^{(n)}&=&0~,
\label{analitycity}
\eea
the infinitesimal isometry transformation acts on it as follows:
\bea
\d_{\x}Q^{(n)}&=&-\Big(\x+{\rm i}\, \r J\Big) \,Q^{(n)}\non\\
&=&
-\Big(\x^{\ha}\cD_{\ha}-\x^{+\hal}\cD^-_{\hal}
+{\rm i}\,\r \big(J^{--}D^{++}
-J^{++}D^{--}
+n J^{+-} \big)\Big)Q^{(n)}~.~~~~~~~~~
\label{SUSYanalyticSuperfield}
\eea
Given two covariantly analytic superfields $Q^{(n)}$ and $Q^{(m)}$, 
their product $Q^{(n)}Q^{(m)}$ is covariantly analytic 
and transforms as $Q^{(n+m)}$.
In addition, the superfield $D^{++}Q^{(n)}$ can be seen 
to be covariantly analytic 
and transform as $Q^{(n+2)}$.

\subsection{Harmonic action principle}

After having introduced various analytic multiplets in ${\rm AdS}^{5|8}\times S^2$,
let us turn to constructing a  supersymmetric action.
It is worth recalling that in the flat global case ($\o=0$), 
the action principle in 5D harmonic superspace
naturally generalizes the original 
4D action rule \cite{HarmonicSuperspace,GIOS} 
and is given by \cite{KuzLin}
\bea
\int \rd^5x \int \rd u\,(\hat{D}^-)^4\,L^{(+4)}\Big|~,
\qquad 
(\hat{D}^-)^4=-{1\over 96}\ve^{\hal\hbe\hga\hde}D^-_{\hal}D^-_{\hbe}D^-_{\hga}D^-_{\hde}~,
\eea
where
$D^-_{\hal} = D^i_{\hal} u^-_i$,  $D^i_{\hal}$ 
are the flat covariant derivatives, and
$L^{(+4)}$ is a real  analytic Lagrangian of harmonic charge $+4$, 
$D^+_{\hal} L^{(+4)}=0$.

We would like to generalize the flat action to the  case of AdS${}^{5|8}$  
using the following ansatz:
\bea
S&=&S_0+a_1S_1+a_2S_2 \non \\
&=&
\int \rd^5x\,e \int \rd u\,\Big{[}(\hat{\cD}^-)^4+a_1
\,\o J^{--}(\hat{\cD}^-)^2+a_2\,(\o J^{--})^2\Big{]}\cL^{(+4)}\big|~,
\label{harmonicAnsatz}
\eea
where $\cL^{(+4)}$ is now covariantly analytic, $\cD^+_{\hal} \cL^{(+4)}=0$, 
\bea
(\hat{\cD}^-)^2\,=\,\cD^{-\hal}\cD^-_{\hal}~, \qquad 
(\hat{\cD}^-)^4\,=\,-{1\over 96}\ve^{\hal\hbe\hga\hde}
\cD^-_{\hal}\cD^-_{\hbe}\cD^-_{\hga}\cD^-_{\hde}~,
\eea
and $a_1,\,a_2$ 
are two constants to be determined.
It is assumed that the above action 
is evaluated in 
Wess-Zumino gauge (\ref{WZ}),
using the bar projection (\ref{projection}),
and as usual $e$
stands for the determinant of the vielbein,
$e=\det(e_{\hm}{}^{\ha})$, with $e_{\hm}{}^{\ha} e_{\ha}{}^{\hn} = \d_{\hm}{}^{\hn}$.

In accordance with the definition of $S$, there are several rules for integration 
by parts which  one can use in practice:
\bea
\int \rd^5x\,e \int \rd u\,\cD_{\hat a} Q^{\hat a} |&=&0~, 
\label{i-b-p-spacetime}\\
\int \rd^5x\,e \int \rd u\, D^{++}Q^{--}|&=& \int \rd^5x\,e \int \rd u\,
  D^{--}Q^{++}|=0~, 
\label{i-b-p-harmonic}\\\
\int \rd^5x\,e \int \rd u\,J\, Q^{(0)} |&=&0~. 
\label{i-b-p-internal}  
\eea
Here $Q^{(0)}$ is a covariantly analytic superfield of harmonic charge 0.

Our aim is to find the constants $a_1,\,a_2$ for which $S$ is invariant under 
the isometry transformations of AdS${}^{5|8}$.  
Let us first compute the variation of $S_0$ under infinitesimal 
isometry transformations. 
Due to (\ref{killing2}), we have 
\bea
\d \Big((\hat{\cD}^-)^4\cL^{(+4)}\Big) =(\hat{\cD}^-)^4 \d \cL^{(+4)}
&=&-(\hat{\cD}^-)^4\Big(\x+{\rm i}\, \r J\Big)\cL^{(+4)} \non \\
=-\Big(\x+
{\rm i} \,\r J+\L^{\hal\hbe}M_{\hal\hbe}\Big)(\hat{\cD}^-)^4\cL^{(+4)}
&=&-\Big(\x+
{\rm i} \,\r J\Big)(\hat{\cD}^-)^4\cL^{(+4)}~.
\eea
Since $ \cL^{(+4)}$ is covariantly analytic, we obtain
\bea
\d_{\x}S_0
&=&-\int \rd^5x\, e \int \rd u\,\Big(\x+
{\rm i} \,\r J
\Big)(\hat{\cD}^-)^4\cL^{(+4)}\Big|\non\\
&=&\int \rd^5x\, e \int \rd u\,\Big{(}\x^{+\hal}\cD^-_{\hal}(\hat{\cD}^-)^4
-\x^{-\hal}{[}\cD^+_{\hal},(\hat{\cD}^-)^4{]}\Big{)}\cL^{(+4)}\Big|~.
\label{varS_0}
\eea
Here we have also used eqs. (\ref{i-b-p-spacetime}) and (\ref{i-b-p-internal}). 

To compute $\cD^-_{\hal}(\hat{\cD}^-)^4$ in (\ref{varS_0}), we observe that
$
\cD^-_{[\hal}\cD^-_{\hbe}\cD^-_{\hga}\cD^-_{\hde}\cD^-_{\hrh]}=0,
$
and then
\bea
0&=&5 \ve^{\hbe\hga\hde\hrh}\cD^-_{[\hal}\cD^-_{\hbe}\cD^-_{\hga}\cD^-_{\hde}\cD^-_{\hrh]}\non\\
&=&
\ve^{\hbe\hga\hde\hrh}\Big{[}
\cD^-_{\hal}\cD^-_{\hbe}\cD^-_{\hga}\cD^-_{\hde}\cD^-_{\hrh}+
\cD^-_{\hrh}\cD^-_{\hal}\cD^-_{\hbe}\cD^-_{\hga}\cD^-_{\hde}+
\cD^-_{\hde}\cD^-_{\hrh}\cD^-_{\hal}\cD^-_{\hbe}\cD^-_{\hga}\non\\
&&~~~~~~~+\cD^-_{\hga}\cD^-_{\hde}\cD^-_{\hrh}\cD^-_{\hal}\cD^-_{\hbe}+
\cD^-_{\hbe}\cD^-_{\hga}\cD^-_{\hde}\cD^-_{\hrh}\cD^-_{\hal}\Big{]}
~.
\eea
Moving $\cD^-_{\hal}$ in each term to the left gives
\bea
\cD^-_{\hal}(\hat{\cD}^-)^4&=&{\o J^{--}\over 120}\,\ve^{\hbe\hga\hde\hrh}\Big{[}
\,4\,M_{\hal\hbe}\cD^-_{\hga}\cD^-_{\hde}\cD^-_{\hrh}
+3\,\cD^-_{\hga}M_{\hal\hbe}\cD^-_{\hde}\cD^-_{\hrh}\non\\
&&~~~~~~~~~~~~~~~~\,+2\,\cD^-_{\hga}\cD^-_{\hde}M_{\hal\hbe}\cD^-_{\hrh}
+\cD^-_{\hga}\cD^-_{\hde}\cD^-_{\hrh}M_{\hal\hbe}\,\Big{]}~.
\eea
This can be further transformed by moving all the Lorentz generators to the right
and factors of $\cD^-_{\hal}$ to the left
using iteratively the algebra of covariant derivatives. 
We end up with 
\bea
\cD^-_{\hal}(\hat{\cD}^-)^4&=&\, \cD^-_{\hal} \Big\{
-{5\over 12}\,\o J^{--}
(\hat{\cD}^-)^2
+3\,(\o J^{--})^2
\Big\}
\non \\
&&-\,\Big({1\over 8}\,\o J^{--}\cD^{-\hbe}(\hat{\cD}^-)^2
-\,{8\over 3}\,(\o J^{--})^2\cD^{-\hbe}
+{1\over 2}\,(\o J^{--})^2\cD^-_{\hga}M^{\hga\hbe}\Big)M_{\hbe\hal}~.~~~~~~
\label{-----}
\eea
The expression in the second line does not contribute when acting on a Lorentz
scalar such as $\cL^{(+4)}$.

To compute $[\cD^+_{\hal},(\hat{\cD}^-)^4]$ in (\ref{varS_0}),
we should iteratively use the algebra of covariant derivatives.
This is an obvious but tedious procedure. The result is:
\bea
{[}\cD^+_{\hal},(\hat{\cD}^-)^4{]}&=&
{1\over 4}\ri \cD_{\hal\hbe}\cD^{-\hbe}(\hat{\cD}^-)^2
+{3\over 8}\o J\cD^-_{\hal}(\hat{\cD}^-)^2
-{1\over 3}\o J^{--}\ri \cD_{\hal\hbe}\cD^{-\hbe}\non\\
&&-\,{13\over 2}\o^2 J^{--} J\cD^-_{\hal}
+{1\over 4}\o J^{+-}\cD^-_{\hal}(\hat{\cD}^-)^2
-5\o^2J^{--}J^{+-}\cD^-_{\hal}\non\\
&&-\,{1\over 8}\o J^{--}(\hat{\cD}^-)^2\cD^+_{\hal}
+{3\over 8}\o J^{--}\cD^-_{\hal}\cD^-_{\hbe}\cD^{+\hbe}
-{3\over 8}\o J^{--}\cD^-_{\hbe}\cD^-_{\hal}\cD^{+\hbe}\non\\
&&+\,{55\over 12}(\o J^{--})^2\cD^+_{\hal}
+\o J^{--}\ri \cD_{\hal\hbe}\cD^-_{\hga}M^{\hbe\hga}
-{3\over 2}\o^2J^{--}J\cD^{-\hbe}M_{\hbe\hal}\non\\
&&+\,2\o^2J^{--}J^{+-}\cD^{-\hbe}M_{\hbe\hal}
-{1\over 4}\o J^{+-}\cD^{-\hbe}(\hat{\cD}^-)^2M_{\hbe\hal}\non\\
&&-\,\o^2J^{--}J^{+-}\cD^-_{\hga}M^{\hbe\hga}M_{\hbe\hal}~.\label{[+,----]}
\eea
Using  the relations (\ref{-----}) and (\ref{[+,----]}), and 
also the integration by parts identities (\ref{i-b-p-spacetime}) and (\ref{i-b-p-internal}), 
variation (\ref{varS_0}) turns into
\bea
\d_{\x}S_0&=&\int \rd^5x\, e \int \rd u\,\x^{+\hal}\Big{[}
-{5\over 12}\,\o J^{--}\cD^-_{\hal}(\hat{\cD}^-)^2
+3\,(\o J^{--})^2\cD^-_{\hal}\,\Big{]}\cL^{(+4)}\Big|\non\\
&&+\int \rd^5x\, e \int \rd u\,\Big{[}-{1\over 4}
(\ri \cD_{\hal\hbe}\x^{-\hbe})\cD^{-\hal}(\hat{\cD}^-)^2
+{3\over 8}\o (J\x^{-\hal})\cD^-_{\hal}(\hat{\cD}^-)^2
\non\\
&&~~~~~~~~~~~~~~~~~~~~~~\,
+{1\over 3}\o J^{--}(\ri \cD_{\hal\hbe}\x^{-\hbe})\cD^{-\hal}
-{13\over 2}\o^2 J^{--} (J\x^{-\hal})\cD^-_{\hal}
\non\\
&&~~~~~~~~~~~~~~~~~~~~~~~
-{1\over 4}\o J^{+-}\x^{-\hal}\cD^-_{\hal}(\hat{\cD}^-)^2
+5\o^2J^{--}J^{+-}\x^{-\hal}\cD^-_{\hal}\,\Big{]}\cL^{(+4)}\Big|
~.~~~~~~
\eea
${}$Finally, it remains to note 
$J\x^-_{\hal}=J^{--}\x^+_{\hal}-J^{+-}\x^-_{\hal}$,  and 
also make use of eq. (\ref{KillingDiracEq}) 
projected to the  minus-harmonics
\be
\ri\cD_{\hal\hbe}\,\x^{-\hbe}\,=\,-{5\over 2}\,\o (J^{--}\x^+_{\hal}-J^{+-}\x^-_{\hal})~.
\ee
As a result, the variation of $S_0$
under the isometry transformations takes the final form:  
\bea
\d_{\x}S_0&=&\int \rd^5x\, e \int \rd u\,\Big{[}\,
-{2\over 3}\,\o J^{--}\x^{+\hal}\cD^-_{\hal}(\hat{\cD}^-)^2
-{8\over 3}\,(\o J^{--})^2\x^{+\hal}\cD^-_{\hal}\non\\
&&~~~~~~~~~~~~~~~~~~\,+{32\over 3}\,\o^2J^{--}J^{+-}\x^{-\hal}
\cD^-_{\hal}\,\Big{]}\cL^{(+4)}\Big|~.
\label{SUSYvarS_0}
\eea

The next step is to compute the variation of the functional $S_1$ 
appearing in our
action  (\ref{harmonicAnsatz}). Here the procedure is the same as for $S_0$.
Varying
\bea
\d \Big((\hat{\cD}^-)^2\cL^{(+4)}\Big) 
=-(\hat{\cD}^-)^2\Big(\x+{\rm i}\, \r J\Big)\cL^{(+4)} 
=-\Big(\x+
{\rm i} \,\r J\Big)(\hat{\cD}^-)^2\cL^{(+4)}~,
\eea
we get
\bea
\d_{\x}S_1
&=&\int \rd^5x\, e \int \rd u\,\Big{(}
\o J^{--}\x^{+\hal}\cD^-_{\hal}(\hat{\cD}^-)^2
-\o J^{--}\x^{+\hal}{[}\cD^+_{\hal},(\hat{\cD}^-)^2{]}\,\Big{)}\cL^{(+4)}\Big|~.~~~~~~
\label{varS_1}
\eea
Using the algebra of covariant derivatives gives 
\bea
{[}\cD^+_{\hal},(\hat{\cD}^-)^2{]}&=&-4\ri\cD_{\hal\hbe}\cD^{-\hbe}-6\o J\cD^-_{\hal}
+12\o J^{+-}\cD^-_{\hal}\non\\
&&-2\o J^{--}\cD^+_{\hal}+8\o J^{+-}\cD^{-\hbe}M_{\hbe\hal}~.
\eea
As a result, the  variation of $S_1$ is
\bea
\d_{\x}S_1&=&\int \rd^5x\, e \int \rd u\,\Big{[}\,
\o J^{--}\x^{+\hal}\cD^-_{\hal}(\hat{\cD}^-)^2
+4(\o J^{--})^2\x^{+\hal}\cD^-_{\hal}\non\\
&&~~~~~~~~~~~~~~~~~~~\,-16\o^2J^{--}
J^{+-}\x^{-\hal}\cD^-_{\hal}\,\Big{]}\cL^{(+4)}\Big|~.
\label{SUSYvarS_1}
\eea
It is seen that
(\ref{SUSYvarS_1}) is proportional to (\ref{SUSYvarS_0}).
Therefore, 
 our ansatz (\ref{harmonicAnsatz})
 leads to the unique supersymmetric action:
$a_1=2/3$ and $a_2=0$.

The supersymmetric action is
\bea
S&=&\int \rd^5x\,e \int \rd u\,\Big\{ 
(\hat{\cD}^-)^4
+{2\over 3}\,\o\,
J^{--}(\hat{\cD}^-)^2 \Big\}
\cL^{(+4)}\Big| ~,\qquad \cD^+_{\hal} \cL^{(+4)}=0~.
\label{harmonicAction}
\eea
This is the main result of this section. 

By construction, the Lagrangian in (\ref{harmonicAction})
is a covariantly analytic superfield of harmonic charge $+4$. 
It should be also chosen to be real with respect to analyticity preserving 
conjugation \cite{HarmonicSuperspace} (see also subsection 
5.1), and then  action (\ref{harmonicAction})
can be seen to be real. Otherwise, $\cL^{(+4)}$ is completely arbitrary. 
Therefore, a great many flat superspace actions \cite{GIOS} 
can be lifted to the AdS superspace. For instance, an off-shell hypermultiplet
can be realized in terms of a covariantly analytic superfield $q^+(z,u)$ and its 
conjugate $\tilde{q}^+(z,u)$, with respect to the anlyticity preserving conjugation. 
To describe its dynamics, one can choose 
\be
 \cL^{(+4)}=-\tilde{q}^+ D^{++} q^+ +\l\, (\tilde{q}^+q^+)^2~,
 \ee
 with $\l$ a coupling constant.

\sect{Projective superspace approach}
\label{sectProjectiveSuperspace}

In the projective superspace approach to $d$-dimensional theories 
with eight supercharges, one deals with superfields that live in 
$\cM^{d|8}\times S^2$, where $\cM^{d|8}$ denotes the conventional superspace,
$d \leq 6$, 
and $S^2$ the two-sphere. Such superfields are required to  (i) be Grassmann analytic, 
i.e. to be annihilated by  one  half of  the supercharges; 
(ii) be holomorphic on an open domain of $S^2$.
The latter requirement is equivalently achieved by considering superfields $\J^{(n)}(z, u^+)$
which are holomorphic 
functions of a single isotwsitor 
$u^{+i} \in {\mathbb C}^2-\{0\}$,  
and have definite degree of homogeneity with respect to $u^+$,  
$\J^{(n)}(z, c \,u^+) =c^n\, \J^{(n)}(z, u^+)$.
The variables $u^{+i}$ can be viewed as  homogeneous coordinates for ${\mathbb C}P^1$.
A second linearly independent isotwistor, $u^{-i}$, is only required
(as a purely auxiliary means, without any intrinsic significance) 
for constructing a supersymmetric  
action which was proposed originally in four dimensions in 
\cite{ProjectiveSuperspace} and then reformulated in \cite{SiegelProjective}
in terms of the projective isotwistor $u^{+i}$. The terminology ``isotwistor''
is due to \cite{Rosly2,RS}.

In the flat global case, 
the 5D $\cN=1$ extension of the 4D $\cN=2$  supersymmetric action \cite{SiegelProjective}
is as follows\footnote{Note that the action given in eq. (B.1) of \cite{KuzConformal5D}
contains a wrong overal factor 
of $\sqrt{-1}$.}
\cite{KuzConformal5D}:
\bea
-{1\over 2\pi}
\oint {u_i^+\rd u^{+i}\over (u^+u^-)^4}\int\rd^5 x\,(\hat{D}^-)^4L^{++}(z,u^+)\big|~,\label{flatSiegelProjectiveAction}
\eea
where
\be
D^+_{\hal}L^{++}(z,u^+)\,=\,0~,\qquad L^{++}(z,c\,u^+)\,=\,c^2\,L^{++}(z,u^+)~,\qquad
c\in \mathbb{C}^*~.
\ee
The action is invariant under arbitrary
projective transformations of the form
\be
(u_i{}^-\,,\,u_i{}^+)~\to~(u_i{}^-\,,\, u_i{}^+ )\,R~,~~~~~~R\,=\,
\left(\begin{array}{cc}a~&0\\ b~&c~\end{array}\right)\,\in\,{\rm GL(2,\mathbb{C})}~.
\label{projectiveGaugeVar}
\ee
This gauge-like  symmetry implies that the action is actually independent of $u^-_i$. 
It can be fixed by imposing, for instance,  the gauge
\bea
u^{+i}\, \sim \,(1,\z)\,=\,\z^i \quad &
\longrightarrow
&\quad 
 u^+_i\,\sim\,(-\z,1)\,=\,\z_i~,\non\\
u^{-i}\,\sim\,(0,-1) \quad & \longrightarrow & \quad u^-_i\,\sim\,(1,0)~,
\label{goodGauge}
\eea
in which the action
(\ref{flatSiegelProjectiveAction}) reduces to the standard 
5D  $\cN=1$ projective superspace action
\cite{KuzLin,KuzConformal5D}. 

\subsection{Projective multiplets}

Here we introduce several off-shell projective multiplets
that are most interesting from the point of view of  model building. 
By definition, a projective superfield $Q^{(n)}(z,u^+)$ 
lives on the anti-de Sitter superspace and depends parametrically 
on a non-vanishing isotwistor $u^{+i}\neq 0$.  
It is defined to be analytic, 
\be
\cD^+_{\hat{\a}} Q^{(n)} =0~, 
\ee
and transform by the rule
\be
\d Q^{(n)} = -\big( \x + {\rm i}\,\r J \big) Q^{(n)}
\label{pr-tr-law}
\ee
under the isometry group. 
We specify $J$ to act on $Q^{(n)}$ as follows
\bea
J \,Q^{(n)}= -\frac{ J^{++}D^{--}Q^{(n)}- nJ^{+-}Q^{(n)} }{(u^+u^-)}~.
\label{J-Q}
\eea
This definition involves an external isotwistor $u^-_i$  subject to the only
requirement 
\be
(u^+u^-) \neq 0~.
\ee
Since $Q^{(n)}$ is independent of $u^-$, it is natural to require $J \,Q^{(n)}$ 
to be independent of $u^-$ as well, that is 
\be
J^{++} \frac{\pa }{\pa u^{+i}}  Q^{(n)}- nJ_{ij} u^{+j} Q^{(n)} 
 =u^+_i J\,Q^{(n)}~.
\ee
Contracting this with $u^{+i}$ gives
\be
J^{++} u^{+i}\frac{\pa }{\pa u^{+i}}  Q^{(n)}= nJ^{++} Q^{(n)} ~.
\ee
Therefore, $Q^{(n)}$ is a homogeneous function of $u^+$ of degree $n$,
\be
Q^{(n)}(z,c\,u^+)\,=\,c^n\,Q^{(n)}(z,u^+)~, \qquad c\in \mathbb{C}^*~.
\label{weight}
\ee
The $Q^{(n)}$ will be  called a projective superfield of weight $n$.

As is obvious, the complex conjugate of an analytic superfield is not analytic.
However, one can introduce a generalized,  analyticity-preserving 
conjugation  \cite{Rosly,HarmonicSuperspace,ProjectiveSuperspace},
$ u^{+i} \to \widetilde{u^{+i}}$,
which is obtained by composing the complex conjugation, 
$u^{+i} \to \overline{u^{+i}}$,
with the antipodal map $\overline{u^{+i} } \to -u^+_i$.
In what follows,  it is called ``smile-conjugation.'' 
It is thus defined to act on the isotwistor $u^+=(u^{+i})$ 
by the rule\footnote{Due to  projective invariance,
$u^{+i} \sim c \,u^{+i}$, the smile-conjugation could be also
defined as $u^+ ~\to ~\tilde{u}^+ = -{\rm i}\, \s_2\, u^+$, instead of 
(\ref{smile}).}
\be
u^+ ~\to ~\tilde{u}^+ = {\rm i}\, \s_2\, u^+~,
\qquad \widetilde{\widetilde{{(u^{+i}})}}= -u^{+i}~,
\label{smile}
\ee
with $\s_2$ the second Pauli matrix.
Its action on the  projective superfields is defined to be 
 \be
Q^{(n)} (u^+) \to \widetilde{Q}^{(n)} (u^+)\equiv \bar{Q}^{(n)}\big(\widetilde{u}^+\big)
~,	\qquad 
\widetilde{\widetilde{Q}}{}^{(n)}
\,=\,(-1)^nQ^{(n)}~.
\label{smileSuperfields}
\ee
It is clear that $\widetilde{Q}^{(n)} (u^+)$ is a homogeneous function of $u^+$ 
of degree $n$, that is  $\widetilde{Q}^{(n)} (c\,u^+) =c^n \,\widetilde{Q}^{(n)} (u^+)$,
with $c\in \mathbb{C}^*$. 
Due to the identity
\be
\widetilde{ {\cD^+_{\hal} Q^{(n)}} }\,=\,(-1)^{\e(Q^{(n)})}\,\cD^{+\hal} \widetilde{Q}{}^{(n)}~,
\ee
the smile-conjugation indeed preserves analyticity.

It is important to note that, in accordance with  (\ref{smileSuperfields}), 
for  an even integer weight, $n=2p$,  one can consistently  
define  real projective superfields $R^{(2p)}$ with respect to the smile-conjugation: 
$\widetilde{R}^{(2p)}=R^{(2p)}$. 

Now, let us show that the smile-conjugation is compatible
with the  superfield transformation law (\ref{pr-tr-law}). 
To evaluate the smile-conjugate of $J\,Q^{(n)}$, eq. (\ref{J-Q}),
we conventionally define the operation of smile-conjugate  for $u^-=(u^{-i})$ 
to be identical to that we have already chosen for the isotwistor $u^+$, that is 
\be
u^- ~\to ~\tilde{u}^- = {\rm i}\, \s_2\, u^-~,
\qquad \widetilde{\widetilde{{(u^{-i}})}}= -u^{-i}~,
\label{smile-}
\ee
We should emphasize that such a definition  is completely 
conventional in the sense that 
the projective superfields
are independent of the isotwistor $u^-$. 
Then it holds
\bea
\widetilde{D^{--}}=D^{--}~,~~~\widetilde{J^{\pm\pm}}\,=\,-J^{\pm\pm}~,~~~
\widetilde{J^{+-}}\,=\,-J^{+-}~,~~~\widetilde{(u^+u^-)}\,=\,(u^+u^-)~.~~~
\eea
This implies
\bea
\widetilde{J\,Q^{(n)}}&=&-\,J\,\widetilde{Q}^{(n)}~,
\eea
and the smile-conjugate of the transformation law (\ref{pr-tr-law}) is 
\bea
\widetilde{{\d Q^{(n)}}}&=&
-(\x+ {\rm i}\,\r J)\,\widetilde{Q}^{(n)}=\d\,\widetilde{Q}^{(n)}~.
\eea
Therefore, the smile-conjugation preserves the superfield transformation  
laws under the isometry group.

As is known, the space ${\mathbb C}P^1$ can be covered
by two charts that are defined in terms of $u^+ = (u^{+\1}, u^{+\2})$
as follows: (i) the north chart on which $u^{+\1} \neq 0$; 
(ii) the south chart on which $u^{+\2} \neq 0$. 
As will be described below, the projective action 
involves the line  integral over a closed contour in ${\mathbb C}P^1$,  
and this contour can be chosen to lie inside one of the coordinate charts.
The latter can be chosen to be the north chart, and that is why our local considerations 
will be mainly restricted to that chart. In the north chart, we can introduce a 
projective invariant complex coordinate $\z$ defined as 
$u^{+i} = u^{+\1}\,(1,\z)$, with $\z= u^{+\2}/u^{+\1}$.
Since $\tilde{u}^{+i} = (u^{+\2}, -u^{+\1})$, the smile-conjugation 
acts as follows:
\be
\z~\to ~ - \frac{1}{\z}~.
\ee

The simplest solution to eq. (\ref{weight})
is the $O(n)$ multiplet defined by eqs.  
(\ref{O(n)1})
and (\ref{O(n)2}).
This multiplet is globally defined on ${\mathbb C}P^1$.
Allowing for singularities at  some points in ${\mathbb C}P^1$
offers the possibility to generate many more interesting supermultiplets.
${}$For example, a charged hypermultiplet is described 
by a weight-one projective superfield $\U^+(u^+)$ being holomorphic
on ${\mathbb C}P^1 - \{N \}$, where the 
North pole is identified
with $u^{+i} \sim (0,1)$. We can represent $\U^+(u^+)$ as
\be
\U^+(u^+)  =u^{+\underline{1}} \, \U^+(u^{+i}/u^{+\underline{1}})
\equiv u^{+\underline{1}} \,\U(\z)~, 
\qquad \U(z,\z) =\sum_{k=0}^{+\infty} \U_k(z) \z^k~.
\ee
Its smile-conjugate  $\widetilde{\U}^+(u^+)$ is holomorphic
on ${\mathbb C}P^1 - \{S \}$, where the 
South pole is identified
with $u^{+i} \sim (1,0)$. We can represent $\widetilde{\U}^+(u^+)$ as
\be
\widetilde{\U}^+(u^+)  =u^{+\underline{2}} \, \widetilde{\U}^+(u^{+i}/u^{+\underline{2}})
\equiv u^{+\underline{2}} \,\widetilde{\U}(\z)~, 
\qquad \widetilde{\U}(z,\z) =\sum_{k=0}^{+ \infty} 
\bar{\U}_k(z) \,
\frac{(-1)^k}{\z^k}~,
\ee
with $\bar{\U}_k(z) $ the complex conjugate of $\U_k(z) $.
To describe an off-shell vector multiplet, one should use 
a real weight-zero projective superfield $V(u^+)$ being holomorphic
on ${\mathbb C}P^1 - \{N \cup S\}$. It can be represented as 
\be
 V(z,\z) =\sum_{k=-\infty}^{+\infty} V_k(z) \z^k~, \qquad \bar{V}_k = (-1)^k V_{-k}~.
 \label{VM}
\ee

\subsection{Projective action principle}
\label{PAP}

Our aim here is 
to find a generalization 
of the flat superspace action (\ref{flatSiegelProjectiveAction}) 
to the  case of  AdS${}^{5|8}$ superspace. 
We start with the following ansatz\footnote{An alternative approach to 
introduce the projective action consists in using a proper generalization
of the procedure given in \cite{Kuzenko}.
The latter allows one to derive the projective action as a singular limit of the harmonic action.}
\bea
S&=&S_0+\b_1\,S_1+\b_2\,S_2\non\\
&=&-{1\over 2\pi}
\oint {u_i^+\rd u^{+i}\over (u^+u^-)^4}
\int\rd^5 x \,e
\Big{[}(\hat{\cD}^-)^4+\b_1\o J^{--}(\hat{\cD}^-)^2+\b_2(\o J^{--})^2\Big{]}\cL^{++}(z,u^+)\Big|~.~~~~~~
\label{projectiveAnsatz}
\eea
Here $\cL^{++}(z,u^+) $ is a covariantly analytic superfield,
$\cD^+_{\hat{\a}} \cL^{++} =0$, which is homogeneous in $u^+_i$ of degree $+2$.
The line integral in (\ref{projectiveAnsatz}) is carried out over a closed contour,
$\g =\{u^+_i(t)\}$,
in the space of $u^+$ variables. 
The integrand in (\ref{projectiveAnsatz}) involves a constant (i.e. time-independent)
isotwistor $u^-_i$ subject to the only condition that $u^+(t)$ and $u^-$ form 
a linearly independent basis at each point of the contour $\g$, that is 
$(u^+ u^-) \neq 0$. 

Our first requirement is that the action 
(\ref{projectiveAnsatz}) be invariant under 
the projective gauge transformations 
(\ref{projectiveGaugeVar}). First of all, it is obvious that 
(\ref{projectiveAnsatz}) 
is invariant under arbitrary scale 
transformations $u^+_i (t) \to c(t)\,u^+_i(t) $, with $c(t) \neq 0$.
It is thus only necessary to analyse projective transformations of $u^-$ of the form
\be
u^-_i~\to~\tilde{u}^-_i\,=\,a(t)\,u^-_i+b(t)\,u^+_i(t)~, \qquad a(t) \neq 0~.
\ee
Since both $u^-$ and $\tilde{u}^-$ should be time independent, 
the coefficients should obey the equations
(using the notation $\dt{f}\equiv \rd f(t)/ \rd t$, for a function $f(t)$):
\bea
\dt{a}=b\,{(\dt{u}^+u^+)\over (u^+u^-)}~, \qquad \dt{b}=-b\,{(\dt{u}^+u^-)\over (u^+u^-)}~.
\label{ode}
\eea
As is obvious, the action (\ref{projectiveAnsatz}) is invariant under 
arbitrary scale transformations $u^-_i \to a(t)\,u^-_i $, with $a \neq 0$.
Therefore, it only remains  to analyse infinitesimal transformations of the form
$\d u^-_i = b(t) u^+_i$, with $b(t)$ obeying the differential equation  (\ref{ode}).
This transformation induces the following variations: 
\be
\d\cD^-_{\hal}\,=\,b\,\cD^+_{\hal}~, \qquad \d J^{--}  \,=\, 2b\, J^{+-}~.
\ee

Let us start by evaluating the variation of $S_0$. Using the condensed 
notation  
\be
{\rm d} \mu^{++}\equiv -{1\over 2\pi}{u_i^+\rd u^{+i}\over (u^+u^-)^4}
=  -{1\over 2\pi}{(\dt{u}^+ u^+) \over (u^+u^-)^4}\,{\rm d} t 
~,
\ee
we obtain
\bea
\d S_0&=&
\oint {\rm d} \mu^{++}
\int\rd^5 x\,e
\,\Big{[}\d(\hat{\cD}^-)^4\Big{]}\cL^{++}\Big|\non\\
&=&-{\ve^{\hal\hbe\hga\hde}\over 96}
\oint
{\rm d} \mu^{++} \,b
\int\rd^5 x\,e
\,\Big{[}
\,3\,\cD^-_{\hal}\cD^-_{\hbe}\{\cD^+_{\hga},\cD^-_{\hde}\}
+2\,\cD^-_{\hal}\{\cD^+_{\hbe},\cD^-_{\hga}\}\cD^-_{\hde}\non\\
&&~~~~~~~~~~~~~~~~~~~~~~~~~~~~~~~~~~~
+\{\cD^+_{\hal},\cD^-_{\hbe}\}\cD^-_{\hga}\cD^-_{\hde}\,\Big{]}\cL^{++}\Big|~.
\label{projectiveActionGaugeVar}
\eea
Now, making use  of the covariant derivatives algebra (\ref{algebra1++}--\ref{algebra2-.2}) 
and the identities 
\begin{subequations}
\bea
{[}J,\cD^+_{\hal}{]}&=&{1\over (u^+u^-)}\left(J^{+-}\cD^+_{\hal}-J^{++}\cD^-_{\hal}\right)~,
\label{[J,+]}
\\
{[}J,\cD^-_{\hal}{]}&=&{1\over (u^+u^-)}\left(J^{--}\cD^+_{\hal}
-J^{+-}\cD^-_{\hal}\right)~,\label{[J,-]}
\eea
\end{subequations}
we can systematically move in (\ref{projectiveActionGaugeVar})
 all space-time derivatives to the left 
(neglecting total space-time derivatives) and the $J$ operator to the right.
 This gives
\bea
\d S_0\,=\,
\oint 
{\rm d}\mu^{++}
\,b
\int\rd^5 x\,e
&\Big{[}&
-{7\over 12}\o J^{+-}(\hat{\cD}^-)^2
-{3\over 8}\o(u^+u^-)(\hat{\cD}^-)^2J
\non\\
&&+{11\over 2}(u^+u^-)\o^2J^{--}J\,\Big{]}\cL^{++}\Big|~.
\label{projectiveGaugeVar2}
\eea
To transform the second and third terms in the square brackets, we should first
recall how $J$ acts on the Lagrangian,
\bea
J\,\cL^{++}&=&{1\over (u^+u^-)}\left(-J^{++}D^{--}\cL^{++}+2J^{+-}\cL^{++}\right)~.
\label{JLprojectiveAction1}
\eea
Since $\cL^{++} $ is a homogeneous function of degree two, we also have 
\bea
{\rd\over \rd t}\cL^{++}&=&{(\dt{u}^+u^-)\over (u^+u^-)}u^{+i}{\pa\over \pa u^{+i}}\cL^{++}
-{(\dt{u}^+u^+)\over (u^+u^-)}D^{--}\cL^{++}\non\\
&=&2{(\dt{u}^+u^-)\over (u^+u^-)}\cL^{++}
-{(\dt{u}^+u^+)\over (u^+u^-)}D^{--}\cL^{++}~.
\label{timeDerivativesProjectiveAction}
\eea
The latter results leads to
\bea
(\dt{u}^+u^+)\,J\,\cL^{++}&=&
J^{++}{\rd\over \rd t}\cL^{++}
- 2{(\dt{u}^+u^-)\over 
(u^+u^-)}J^{++}\cL^{++}
+2{(\dt{u}^+u^+)
\over (u^+u^-)}J^{+-}\cL^{++}~.
\label{JLprojectiveAction2}
\eea
One more technical observation, 
\bea
{\rd\over \rd t}J^{++}&=&2{(\dt{u}^+u^-)\over (u^+u^-)}
J^{++}-2{(\dt{u}^+u^+)\over (u^+u^-)}J^{+-}~,
\label{JLprojectiveAction22}
\eea
allows us to obtain the following identity:
\bea
b\,{(\dt{u}^+u^+)\over (u^+u^-)^3}\,J\,\cL^{++}&=&{\rd\over \rd t}
\left[{b\,J^{++}\over (u^+u^-)^3}\,\cL^{++}\right]+
4b\,{(\dt{u}^+u^+)\over (u^+u^-)^4}J^{+-}\cL^{++}~.
\eea
Then (\ref{projectiveGaugeVar2}) becomes
\bea
\d S_0
&=&
\oint 
{\rm d}\mu^{++}\,b
\int\rd^5 x\,e
\Big{[}
-{25\over 12}\,\o J^{+-}(\hat{\cD}^-)^2
+22\,\o^2J^{--}J^{+-}
\Big{]}
\cL^{++}\Big|
~.~~~
\label{projectiveGaugeVar23}
\eea
Using the same procedure, 
for $\d S_1$ and $\d S_2$ we find
\begin{subequations}
\bea
\d S_1&=&
\oint 
{\rm d}\mu^{++}\,b
\int\rd^5 x\,e
\Big{[}\,
2\,\o J^{+-}(\hat{\cD}^-)^2
+48\,\o^2J^{--}J^{+-}
\Big{]}
\cL^{++}\Big|
~,~~\\
\d S_2&=&
\oint 
{\rm d}\mu^{++}\,b
\int\rd^5 x\,e\,
4\,\o^2J^{--}J^{+-}\cL^{++}\Big|
~.
\eea
\end{subequations}
The relations obtained show that 
the requirement of projective invariance, 
$\d S = \d S_0+\b_1\,\d S_1+\b_2\, \d S_2 =0$, 
uniquely fixes the coefficients in in (\ref{projectiveAnsatz}) as follows:
$\b_1=25/24$ and $\b_2=-18$.
We end up with the projective-invariant action 
\bea
S&=&-{1\over 2\pi}\oint {u_i^+\rd u^{+i}\over (u^+u^-)^4}
\int\rd^5 x\,e\,\Big{[}
(\hat{\cD}^-)^4\,+{25\over 24}\,\o J^{--}(\hat{\cD}^-)^2-18\,(\o J^{--})^2\Big{]}\cL^{++}\Big|~.
~~~
\label{projectiveAction}
\eea

Now,  we are going  to demonstrate that (\ref{projectiveAction})
is  supersymmetric, that is this action is invariant under the isometry group 
of ${\rm AdS}^{5|8}$.  
This requires us to carry out calculations that are very similar to those
presented in section 4 for the harmonic case. But there are two technical 
features being specific for the projective case: (i) unlike the harmonic case, 
we have $(u^+u^-)\neq1$ in general, and therefore it is necessary to keep track 
of the factors of $(u^+u^-)$; (ii) unlike the harmonic superspace identity
(\ref{i-b-p-internal}), in general we have $\oint {\rm d} \mu^{++}J \,Q^{--} \neq 0$.
In all variations involving the U(1) generator $J$,
we will systematically move  
$J$'s  to the right 
to hit the Lagrangian $\cL^{++}$, 
so that
eqs. 
(\ref{JLprojectiveAction1}) and  (\ref{JLprojectiveAction2}) 
can be applied.

We start  by computing the variation of $S_0$ 
under the infinitesimal isometry transformation.
Making use of 
\bea
\d \Big((\hat{\cD}^-)^4\cL^{++}\Big) 
=-(\hat{\cD}^-)^4\Big(\x+{\rm i}\, \r J\Big)\cL^{++} 
=-\Big(\x+
{\rm i} \,\r J\Big)(\hat{\cD}^-)^4\cL^{++}
\eea
gives
\bea
\d_{\x}S_0
&=& \oint {\rm d}\mu^{++}
\int\rd^5x\,e
\Big{[}\,
{1\over(u^+u^-)}\Big(\x^{+\hal}\cD^-_{\hal}(\hat{\cD}^-)^4
-\x^{-\hal}{[}\cD^+_{\hal},(\hat{\cD}^-)^4{]}\Big)
\non\\
&&~~~~~~~~~~~~~~~~~~~~~~
-{\rm i}\,\r\Big({[}J,(\hat{\cD}^-)^4{]}+(\hat{\cD}^-)^4 J \Big)
\Big{]}
\cL^{++}\Big|
~.
\label{susyVarProjectiveAction}
\eea
To evaluate $\cD^-_{\hal}(\hat{\cD}^-)^4\cL^{++}$, we note 
that eq. (\ref{-----}) holds even if $(u^+u^-)\neq1$, 
since  in the derivation of 
(\ref{-----}) we only used eq. (\ref{algebra1--}) and 
the commutation relations of the Lorentz generator $M_{\hal\hbe}$ with 
the covariant derivatives, 
and both results are clearly not affected by the normalization 
of $(u^+u^-)$.
Therefore, for  the first term on the right of (\ref{susyVarProjectiveAction})
we have
\bea
\cD^-_{\hal}(\hat{\cD}^-)^4\cL^{++}&=&
\Big{[}-{5\over 12}\,\o J^{--}\cD^-_{\hal}(\hat{\cD}^-)^2
+3\,(\o J^{--})^2\cD^-_{\hal}\,\Big{]}\cL^{++}~.~~~~~~
\eea

${}$For the operator ${[}\cD^+_{\hal},(\hat{\cD}^-)^4{]}$, which appears 
in  (\ref{susyVarProjectiveAction}),
we have derived eq. (\ref{[+,----]}) in the harmonic case. 
Now, in evaluating the second term on the right of (\ref{susyVarProjectiveAction}), 
we should take care of the factors  of $(u^+u^-)$, as well as to move 
the U(1) generator $J$ to the right. 
This gives
\bea
&& {[}\cD^+_{\hal},(\hat{\cD}^-)^4{]} =
{1\over 4}(u^+u^-)\ri \cD_{\hal\hbe}\cD^{-\hbe}(\hat{\cD}^-)^2
+{3\over 8}(u^+u^-)\o \cD^-_{\hal}(\hat{\cD}^-)^2J 
-{7\over 8}\o J^{+-}\cD^-_{\hal}(\hat{\cD}^-)^2~~~~~~~
\non\\
&&
~~-{11\over 6}(u^+u^-)\o J^{--}\ri \cD_{\hal\hbe}\cD^{-\hbe}
-{17\over 4}(u^+u^-)\o^2 J^{--} \cD^-_{\hal}J
+{33\over 4}\o^2J^{--}J^{+-}\cD^-_{\hal}
+\cdots~,~~~~~~
\label{[+,----]projective}
\eea
where the dots  denote those terms which 
do not contribute when acting on 
 Lorentz scalar and analytic superfields
such as 
the Lagrangian 
$\cL^{++}$.
Inserting (\ref{[+,----]projective}) into $\d S_0$, 
one can get read of the terms with vector covariant derivatives 
by taking into account the integration by parts rule
(\ref{i-b-p-spacetime}) and 
\bea
\ri\cD_{\hal\hbe}\x^{-\hbe}&=&-{5\o\over 2(u^+u^-)}(J^{--}\x^+_{\hal}-J^{+-}\x^-_{\hal})~.
\eea

To evaluate the contributions to $\d S_0$ which contain $J\cL^{++}$, 
we note that 
eqs. (\ref{JLprojectiveAction2})  and (\ref{JLprojectiveAction22})
imply
\bea
\frac{(\dt{u}^+u^+)}{(u^+u^-)^4}
\,J\,\cL^{++}&=&
\frac{\rm d}{{\rm d}t} \Big( \frac{ J^{++}\cL^{++} }{(u^+u^-)^4} \Big)
+4 \frac{(\dt{u}^+u^+)}{(u^+u^-)^5} \,J^{+-} \cL^{++}~.
\eea
The latter  observation tells us
\bea
\oint {\rm d}\mu^{++}
\,\cO(u^-)\,J\,\cL^{++}\Big|&=&
4\oint {\rm d}\mu^{++}
\,{J^{+-}\over (u^+u^-)}\,\cO(u^-)\,\cL^{++}\Big|~,\label{J_on_Projective_Lagrangian}
\eea
for any operator $\cO(u^-)$ independent of $u^+$.  
It follows that
\bea
- \oint {\rm d}\mu^{++}\int \rd^5x\,e\,
{[}\cD^+_{\hal},(\hat{\cD}^-)^4{]}\cL^{++}\Big|&=&
\frac{5}{2}\oint {\rm d}\mu^{++}
\int \rd^5x\,e\,
\frac{\o J^{--}}{(u^+u^-)}
\Big[-\frac{1}{4}
\x^{+\hal}\cD^-_{\hal}(\hat{\cD}^-)^2
\non\\
&&
+\frac{11}{6}\o J^{--}
\x^{+\hal}\cD^-_{\hal}
+
\frac{5}{3}\o J^{+-}\,\x^{-\hal}\cD^-_{\hal}\Big]\cL^{++}\Big|
~.~~~~~~~~
\eea

Completely similar considerations, 
using also $\cD_{\hal\hbe}\,\r=0$ (\ref{covConst}), give
\bea
-\oint {\rm d}\mu^{++}
\int \rd^5x\,e\,
\r{[}J,(\hat{\cD}^-)^4{]}\cL^{++}\big|&=&
\oint {\rm d}\mu^{++}
\int \rd^5x\,e\,
\frac{\r \,J^{+-}}{(u^+u^-)} \Big[
\,4(\hat{\cD}^-)^4
\non\\
&&
+\,\frac{25}{12} \,\o J^{--}(\hat{\cD}^-)^2
-22\,(\o J^{--})^2
\Big]\cL^{++}\Big|~.
\eea
As a result, the variation  
$\d_\x S_0$ can be represented in the form
\bea
\d_{\x}S_0&=&
\oint {\rm d}\mu^{++}
\int \rd^5x\,e
\Big{[}
-{25\over24}{\o J^{--}\over (u^+u^-)}\,\x^{+\hal}\cD^-_{\hal}(\hat{\cD}^-)^2
+{91\over 12}{(\o J^{--})^2\over (u^+u^-)}\,\x^{+\hal}\cD^-_{\hal}
\non\\
&&
+\,{25\over 6}{\o^2 J^{--}J^{+-}\over (u^+u^-)}\,\x^{-\hal}\cD^-_{\hal}
+{25\over 12}{\o J^{--}J^{+-}\over (u^+u^-)}\,\r(\hat{\cD}^-)^2
-22\,{(\o J^{--})^2J^{+-}\over (u^+u^-)}\,\r
\,\Big{]}\cL^{++}\Big|~.~~~~~~~~
\eea

The variations $\d_{\x}S_1$ and $\d_{\x}S_2$ can be computed 
by similar means. The results are:
\begin{subequations}
\bea
\d_{\x}S_1&=&
\oint {\rm d}\mu^{++}
\int \rd^5x\,e
\Big{[}\,
{\o J^{--}\over (u^+u^-)}\,\x^{+\hal}\cD^-_{\hal}(\hat{\cD}^-)^2
+10\,{(\o J^{--})^2\over (u^+u^-)}\,\x^{+\hal}\cD^-_{\hal}
\non\\
&&-\,4\,{\o^2 J^{--}J^{+-}\over (u^+u^-)}\,\x^{-\hal}\cD^-_{\hal}
-2\,{\o J^{--}J^{+-}\over (u^+u^-)}\,\r(\hat{\cD}^-)^2
-48\,{(\o J^{--})^2J^{+-}\over (u^+u^-)}\,\r
\,\Big{]}\cL^{++}\Big|~,~~~~~~~\\
\d_{\x}S_2&=&
\oint {\rm d}\mu^{++}
\int \rd^5x\,e
\Big{[}\,
{(\o J^{--})^2\over (u^+u^-)}\,\x^{+\hal}\cD^-_{\hal}
-4\,{(\o J^{--})^2J^{+-}\over (u^+u^-)}\,\r
\,\Big{]}\cL^{++}\Big|~.
\eea
\end{subequations}
Collecting all the results obtained, we conclude
\bea
\d_{\x}S&=&0~,
\eea
and therefore the action (\ref{projectiveAction}) is supersymmetric.
Actually, it proves to be the only supersymmetric action 
in the family (\ref{projectiveAnsatz}). 
It is quite remarkable that projective invariance implies supersymmetry 
and vice versa.

\sect{Dynamical systems in projective superspace}

In this section we study in more detail the projective multiplets
and then consider several important supersymmetric theories.
To simplify the analysis, it is useful to choose 
the projective gauge 
$u^-_{\2}=0$. 
Without loss of generality, one can also 
work in a representation of the algebra in which $J^{\1\1}
=J^{\2\2}=0$,
and hence 
$J^{--}=0$.

\subsection{Projective multiplets revisited}

In each of the two coordinate charts for ${\mathbb C}P^1$, 
one can describe the projective multiplets by superfields invariant under
the projective transformation (\ref{weight}). Let us restrict our consideration 
to the north chart. Given a complex projective multiplet of weight $n$,
$Q^{(n)}(u^+)$, it can be equivalently described by a holomorphic
function $Q^{[n]} (\z) $ defined as follows:
\be
Q^{(n)}(u^{+i})=(u^{+\1})^nQ^{[n]}(\z)~, \qquad 
Q^{[n]}(\z)\equiv Q^{(n)}(1,\z)~.
\ee
Here $Q^{[n]}(\z)$ is clearly invariant under  (\ref{weight}).
${}$For the smile-conjugate of $Q^{(n)}(u^+)$, we get
\be
\widetilde{Q}^{(n)}(u^{+i})=(u^{+\2})^n\widetilde{Q}^{[n]}(\z)~, \qquad 
\widetilde{Q}^{[n]}(\z) ={\bar Q}^{[n]}(-1/\z)~.
\ee
Given a real projective multiplet $R^{(2p)}(u^+)$, with respect to the smile-coinjugation,
it can be represented 
\be
R^{(2p)}(u^+)=(\ri u^{+\1}u^{+\2})^pR^{[2p]}(\z)~, \qquad 
\widetilde{R}^{[2p]}(\z) \equiv \widetilde{R}^{[2p]}(-1/\z) = R^{[2p]}(\z)~.
\ee

The most general form for $Q^{[n]}(z,\z)$ is 
\be
Q^{[n]}(z,\z)=\sum_{k=-\infty}^{+\infty} Q^{[n]}_{k}(z)\z^k~.
\label{Q[]}
\ee
In the projective  gauge chosen ($u^-_{\2}=0$,  $J^{\1\1}=J^{\2\2}=0$),
the action of the operator $J$ on our superfield becomes
\bea
J \,Q^{[n]}(u^+)&=& -\frac{1}{(u^+u^-)}\left( J^{++}D^{--}- nJ^{+-}\right)Q^{[n]}(u^+)
=
J^{\1\2}\left(n-2u^{+\2}{\pa\over\pa u^{+\2}}\right)Q^{[n]}(u^+)~.
\non
\eea
Then, since the isotwistor $u^{+i}$ is neutral under the action of $J$, 
it holds
\bea
(u^{+\1})^n\,J \,Q^{{[n]}}(\z)=J \,Q^{(n)}(u^{+})&=&
J^{\1\2}\left(n(u^{+\1})^nQ^{{[n]}}(\z)-2u^{+\2}{\pa\over\pa u^{+\2}}
\Big((u^{+\1})^nQ^{{[n]}}(\z)\Big)\right)\non\\
&=&(u^{+\1})^nJ^{\1\2}\Big(nQ^{{[n]}}(\z)-2\z{\pa\over\pa \z}Q^{{[n]}}(\z)\Big)~,
\non
\eea
and therefore
\bea
J \,Q^{[n]}(z,\z)&=&J^{\1}{}_{\1}\left(2\z{\pa\over\pa \z}-n\right)Q^{{[n]}}(z,\z)~,
\qquad J\,Q^{[n]}_k(z)=(2k-n)J^{\1}{}_{\1}Q^{[n]}_k(z)~.~~~
\eea

In the case of  a real 
superfield $R^{(2p)}(z,u^+)=(\ri u^{+\1}u^{+\2})^p R^{[2p]}(z,\z)$,
we have for $R^{[2p]}$ 
\bea
R^{[2p]}(z,\z)=\sum_{k=-\infty}^{+\infty} R^{[2p]}_{k}(z)\z^k~,
\qquad \bar{R}^{{[}2p{]}}_k=(-1)^kR^{{[}2p{]}}_{-k}~.
\label{R[]}
\eea
The operator $J$ is represented as follows:
\bea
J \,R^{[2p]}(z,\z)&=&2J^{\1}{}_{\1}\z{\pa\over\pa \z}R^{{[2p]}}(z,\z)~,
\qquad J\,R^{[2p]}_k(z)=2kJ^{\1}{}_{\1}R^{[2p]}_k(z)~.
\eea

Let us analyse the implications of the analyticity condition,
$\cD^+_{\hat{\a}} Q^{(n)} (u^+)=0$. 
It is useful to change the representation for the projective superfields,
$Q^{(n)} (u^+) \to Q^{[n]} (\z)$. We then have
$\cD^+_{\hal}Q^{[n]}(\z)=-u^{+\1}(\z\cD^{\1}_{\hal}-\cD^{\2}_{\hal})Q^{[n]}(\z)$,
and therefore  the analyticity condition is equivalent to
\be
\cD^{\2}_{\hal}Q^{[n]}(\z) =\z\cD^{\1}_{\hal}Q^{[n]}(\z)~.
\label{4cac1}
\ee
${}$For the component superfields $Q^{[n]}_k(z)$, this implies
\bea
\cD^{\2}_{\hat \a}Q^{[n]}_{k}=\cD^{\1}_{\hat \a}Q^{[n]}_{k-1}~.
\label{4cac2}
\eea
It is natural to think of $\cD^{\1}_{\hat \a}$ and $\cD^{\2}_{\hat \a}$ as the covariant derivatives 
associated with two 5D Dirac spinor coordinates, $\q^{\hat \a}_{\1}$ and their conjugates
 $\q^{\hat \a}_{\2}$. It then follows from (\ref{4cac1}) that the dependence of $Q^{[n]}(\z) $ 
 on $\q^{\hat \a}_{\2}$ is completely determined by the dependence of $Q^{[n]}(\z) $ 
 on $\q^{\hat \a}_{\1}$.
 
 Suppose that the expansion of $Q^{[n]}(\z) $ in powers of $\z$ terminates from below
 \be
Q^{[n]}(z,\z)=\sum_{k=L}^{+\infty} Q^{[n]}_{k}(z)\z^k~.
\label{Q[]2}
\ee
Then, eq. (\ref{4cac2}) tells us that the two lowest components of $Q^{[n]}$ are constrained 
as follows:
\bea
\cD^{\2}_{\hat \a}Q^{[n]}_{L}&=&0~, \non \\
(\hat{\cD}^{\2})^2
Q^{[n]}_{L+1}&=&
12 \o\, J \,Q^{[n]}_{L}~, 
\eea
where
\be
(\hat{\cD}^i)^2 = \cD^{i \hat \a} \cD_{\hat \a}^i~, 
\qquad i= \1 , \2~.
\ee 
Therefore, $Q^{[n]}_{L}$ is a five-dimensional 
{\it chiral superfield}, while $Q^{[n]}_{L+1}$ a 
{\it complex linear  superfield}.
The union of $Q^{[n]}_{L}$ and $Q^{[n]}_{L+1}$ forms a 5D analogue 
of the famous chiral-nonminimal doublet in 4D supersymmetry
\cite{DG}.
 
Given a real $O(2)$ multiplet $H^{(2)}(z,u^+)$, we can represent 
$H^{[2]}(z,\z) $ in the form 
\be
H^{[2]}(z,\z) =\frac{1}{\z} \F(z) + G(z) -\z {\bar \F} (z)~,
\ee
where $\F$ is  a 
five-dimensional 
{\it chiral superfield}, and $G$ a 
{\it real linear  superfield},
\bea
\cD^{\2}_{\hat \a}\F&=&0~, \non \\
(\hat{\cD}^{\2})^2
G&=&0~,\qquad {\bar G}=G~.
\eea

If the expansion of $Q^{[n]}(\z) $ in powers of $\z$ terminates from above, 
 \be
Q^{[n]}(z,\z)=\sum_{k=-\infty}^{L} Q^{[n]}_{k}(z)\z^k~.
\label{Q[]3}
\ee
then eq. (\ref{4cac2}) implies 
that the two highest components of $Q^{[n]}$ are constrained 
as follows:
\bea
\cD^{\1}_{\hat \a}Q^{[n]}_{L}&=&0~, \non \\
(\hat{\cD}^{\1})^2
Q^{[n]}_{L-1}&=&
-12 \o\, J \,Q^{[n]}_{L}~, 
\eea
Therefore, $Q^{[n]}_{L}$ is a 
five-dimensional 
{\it antichiral superfield},
while $Q^{[n]}_{L-1}$ a 
{\it complex antilinear  superfield}.

${}$For further analysis, it is useful to 
switch from the 5D four-component spinor notation to the 4D two-component
one by representing
\bea
\cD^i_{\hat \a}  
&=& \left(
\begin{array}{c}
\cD_\a^i \\
{\bar \cD}^{\ad i}    
\end{array}
\right)\,=\,
\left(
\begin{array}{c}
\cD_\a^i \\
\ve^{ij}{\bar \cD}^{\ad}_j    
\end{array}
\right)~.
\non
\eea
In such a notation, 
the algebra of covariant derivatives 
(\ref{algebra1}--\ref{algebra3}) 
takes the form
\begin{subequations}
\bea
&\{\cD_{\a}^i,\cD_{\b}^j\}\,=\,2\ve^{ij}\ve_{\a\b}\cD_{5}
-3\o\ve^{ij}\ve_{\a\b}J-4\o J^{ij}M_{\a\b}~,
\label{algebra1.1_4D}\\
&\{\cD_{\a}^i,\cDB_j^\bd\}\,=\,-2{\rm i}\d^{i}_{j}(\s^a)_\a{}^\bd\cD_{a}
-4\o J^i_{~j}M_{\a}^{~\bd}~,
\label{algebra1.2_4D}\\
&\{\cDB^{\ad}_i,\cDB^{\bd}_j\}\,=\,-2\ve_{ij}\ve^{\ad\bd}\cD_{5}
-3\o\ve_{ij}\ve^{\ad\bd}J-4\o J_{ij}M^{\ad\bd}~,
\label{algebra1.3_4D}\\
&{[} \cD_{a}, \cD_{\a}^i {]}
\,=\,-{\ri\over 2}\o J^{ij}(\s_{a})_{\a\bd}\cDB_j^{\bd}~,~~~
{[} \cD_{a}, \cDB^{\ad}_i {]}
\,=\,{\ri\over 2}\o J_{ij}(\tilde{\s}_{a})^{\ad\b}\cD^j_{\b}~,
\label{algebra2.1_4D}\\
&{[} \cD_{5}, \cD_{\a}^i {]}
\,=\,{1\over 2}\o J^i_{~j}\cD^j_{\a}~,~~~
{[} \cD_{5}, \cDB^{\ad}_i {]}
\,=\,{1\over 2}\o J_i^{~j}\cDB_j^{\bd}~,
\label{algebra2.2_4D}\\
&{[} \cD_{a},\cD_{b} {]}\,=\,-\o^2 J^2 M_{ab}~,~~~
{[} \cD_{a},\cD_{5} {]}\,=\,-\o^2 J^2 M_{a5}~.
\label{algebra3_4D}
\eea
\end{subequations}

In the two-component spinor notation, the analyticity condition
$\cD^+_{\hal}Q^{[n]}(\z)=0$
is equivalent to 
\be
\cD^{\2}_{\a}Q^{[n]}(\z)\,=\,\z\,\cD^{\1}_{\a}Q^{[n]}(\z)~, \qquad 
\cDB_{\2}^{\ad}Q^{[n]}(\z)\,=\,-{1\over \z}\,\cDB_{\1}^{\ad}Q^{[n]}(\z)~.
\label{ancon22}
\ee
${}$For the component superfields $Q^{[n]}_k(z)$, this implies
\bea
\cD^{\2}_{\a}Q^{[n]}_{k}=\cD^{\1}_{\a}Q^{[n]}_{k-1}~,
\qquad \cDB_{\2}^{\ad}Q^{[n]}_{k}=-\cDB_{\1}^{\ad}Q^{[n]}_{k+1}~.
\label{an4D}
\eea
By analogy with 
the flat case, these constraints indicate an interesting interpretation.
Let us introduce two sets of spinor derivatives,
$(\cD^{\1}_\a, {\bar \cD}^\bd_{\1})$ and $(\cD^{\2}_\a, {\bar \cD}^\bd_{\2})$ 
which can be viewed as the covariant derivatives corresponding 
to two different sets of Grassmann variables $\Q_{\1}$  and $\Q_{\2}$.
Then, the above constraints imply that the dependence 
of the projective superfields on $\Q_{\2}$ 
is uniquely determined in terms of their dependence  on $\Q_{\1}$. 
Unlike the flat case, such an interpretation is somewhat limited 
in the sense that one can not consistently switch off the variables $\Q_{\2}$ 
(what would be necessary for reducing the multiplets to 4D $\cN=1$ superfields). 
It follows from the algebra of covariant derivatives, 
specifically from eq. (\ref{algebra2.1_4D}), that 
$ [ \cD_{a}, \cD_{\a}^{\1} ]
=-{\ri\over 2}\o J^{\1\2}(\s_{a})_{\a\bd}\cDB_{\2}^{\bd}$ and 
$[ \cD_{a}, \cDB^{\ad}_{\1}]
=-{\ri\over 2}\o J^{\1\2}(\tilde{\s}_{a})^{\ad\b}\cD^{\2}_{\b}$, 
and therefore the commutation relations mix all the spinor derivatives.
This is an important difference between the flat and curved  cases.

The constraints (\ref{an4D}) simplify if  the series  in  (\ref{Q[]}) 
or (\ref{R[]}) is bounded from below (above).
Consider
a real $O(2n)$ multiplet $H^{(2n)}(z,u^+)$. 
In accordance with the above general consideration, 
it can be  described by the superfield 
$H^{{[}2n{]}}(z,\z)$ 
which is defined by 
$
H^{(2n)}(z,u^+)=(\ri u^{+\1}u^{+\2})^nH^{{[}2n{]}}(z,\z),
$
and can be represented in the form
\bea
H^{{[}2n{]}}(z,\z)=\sum_{k=-n}^{+n} H^{{[}2n{]}}_{k}(z)\z^k~,
~~~~~~\bar{H}^{{[}2n{]}}_k=(-1)^kH^{{[}2n{]}}_{-k}~.
\eea
The analyticity constraints (\ref{an4D}) imply 
that the two lowest component superfields are constrained by
\bea
\cDB_{\1}^\ad H^{{[}2n{]}}_{-n}&=&0~, \non\\
({\cDB}_{\1})^2H^{{[}2n{]}}_{-n+1}&=&
-(4\cD_5+6\,\o J)H^{{[}2n{]}}_{-n}
=-(4\cD_5-12n\,\o J^{\1}{}_{\1})H^{{[}2n{]}}_{-n}~,
\label{O(n)-4D}
\eea
where we have defined
\bea
(\cD^i)^2\,\equiv\,\cD^{i\a}\cD^i_\a~,~~~~~~(\cDB_i)^2\,\equiv\,\cDB_{i\ad}\cDB_i^\ad~.
\eea

Consider an arctic multiplet of weight $n\geq 0$, $\U^{(n)}(u^+)$, 
defined to be holomorphic
on ${\mathbb C}P^1 - \{N \}$.
It can be represented as
\bea
\U^{(n)}(z,u^+)=(u^{+\1})^n\,\U^{[n]}(z,\z)~,
\qquad 
\U^{[n]}(z,\z)=\sum_{k=0}^{+\infty}\U^{[n]}_k(z)\z^k~.
\label{n-arctic-1}
\eea
Then the constraints on the two lowest components superfields are
 \bea
\cDB_{\1}^\ad \U^{[n]}_0&=&0~,\non\\
({\cDB}_{\1})^2\U^{[n]}_1&=&-(4\cD_5+6\,\o J)\U^{[n]}_0
\,=\,-(4\cD_5-6n\,\o J^{\1}{}_{\1})\U^{[n]}_0~.
\label{n-arctic-2}
\eea
In the flat superspace limit, $\o \to 0$, 
the constraints (\ref{O(n)-4D}) and (\ref{n-arctic-2}) reduce to those 
given in \cite{KuzLin}.

\subsection{Projective action}

Here we turn to a more detailed analysis of the projective action
(\ref{projectiveAction}).
In the projective  gauge  ($u^-_{\2}=0$,  $J^{\1\1}=J^{\2\2}=0$)
used throughout this section,
we have 
$J^{--}=0$,
and therefore the  projective action simplifies 
\bea
S&=&-{1\over 2\pi}\oint {u_i^+\rd u^{+i}\over (u^+u^-)^4}
\int\rd^5 x\,e\,(\hat{\cD}^-)^4\,\cL^{++}\Big|~.
\label{1projectiveActionN=1}
\eea
Of course, the Lagrangian $\cL^{++}$ should be real with respect to the smile conjugation,
and can be represented as 
\be
\cL^{++}(z,u^+)=\ri u^{+\1}u^{+\2}\cL(z,\z)~.
\ee
Then, the  action turns into 
\bea
S&=&
-{1\over 32}\oint {\rd\z \over 2\pi\ri}
\int\rd^5 x\,e\,\z\,
(\hat{\cD}^{\1})^2(\hat{\cD}^{\1})^2\cL(z,\z)\Big|~,
\label{2projectiveActionN=1}
\eea
where we have 
taken into account the fact that 
$\{ \cD^{\1}_{\hal}, \cD^{\1}_{\hbe}\}=0$ 
in the projective gauge, 
and also 
made use of the identity 
$\ve^{\hal\hbe\hga\hde}
=\big(\ve^{\hal\hbe}\ve^{\hga\hde}
+\ve^{\hal\hga}\ve^{\hde\hbe}+\ve^{\hal\hde}\ve^{\hbe\hga}\big)$.
Using the relation
\bea
(\hat{\cD^{\2}})^2Q^{[n]}&=&\z^2(\hat{\cD^{\1}})^2Q^{[n]}+12\o\z \,JQ^{[n]}~,
\eea
we can express action (\ref{2projectiveActionN=1}) in the equivalent forms
\begin{subequations}
\bea
S&=&
-{1\over 32}\oint {\rd\z \over 2\pi\ri \z}
\int\rd^5 x\,e\,
(\hat{\cD}^{\1})^2\Big((\hat{\cD}^{\2})^2-\z\,12\o J\Big)\cL(z,\z)\Big|~,
\label{3aprojectiveActionN=1}
\eea
and
\bea
S&=&
-{1\over 32}\oint {\rd\z \over 2\pi\ri\z}
\int\rd^5 x\,e\,
(\hat{\cD}^{\2})^2\Big((\hat{\cD}^{\1})^2+{1\over\z}\,12\o J\Big)\cL(z,\z)\Big|~,
\label{3bprojectiveActionN=1}
\eea
\end{subequations}
where we have used the identities
\bea
{[}\cD_{\hal}^{\1},(\hat{\cD}^{\2})^2{]}&=&
-18\o J^{\1}_{~\1}\cD^{\2}_{\hal}+4\ri\cD_{\hal\hbe}\cD^{\2\hbe}
+6\o\cD^{\2}_{\hal}J-8\o J^{\1}_{~\1}\cD^{\2\hbe}M_{\hal\hbe}~,\\
{[}(\hat{\cD}^{\1})^2,(\hat{\cDB}^{\2})^2{]}&=&
\,4\ri\cD^{\hal\hbe}{[}\cD^{\1}_{\hal},\cD^{\2}_{\hbe}{]}
+(6\o{[}\cD^{\1}_{\hal},\cD^{\2}_{\hbe}{]}+96\o^2J^{\1}{}_{\1})J
-8\o J^{\1}_{~\1}{[}\cD^{\1}_{\hal},\cD^{\2}_{\hbe}{]}M^{\hal\hbe}~.~~~~~~~
\eea
Then we can represent the action in the form
\bea
S=
-{1\over 32}\oint {\rd\z \over 2\pi\ri \z}
\int\rd^5 x\,e\,\Big(
\Big\{(\hat{\cD}^{\1})^2,(\hat{\cD}^{\2})^2\Big\}+24\o J^{\1}_{~\1}\Big[ \z (\hat{\cD}^{\1})^2
+{1\over \z} (\hat{\cD}^{\2})^2 \Big]\Big)\cL(z,\z)\Big|~~~~
\label{3cprojectiveActionN=1}
\eea
which makes manifest the reality of $S$ with respect to the smile-conjugation.

It can be seen from the above relations that there exists
a natural ``gauge freedom'' in the choice of $\cL^{++}$.
It occurs in the three incarnations:  
\bea
\cL^{++} ~& \to~&~ \cL^{++} +\L^{++}+ \widetilde{\L}^{++}~, 
\label{L++}\\
\cL^{++} ~& \to~&~ \cL^{++} + {\rm i}\,J^{++} \big( \L+ \widetilde{\L} \big) ~,
\label{L}\\
\cL^{++} ~& \to~&~ \cL^{++} + H^{++} ~,
\label{H++}
\eea
with $\L^{++}$ and $\L$ arctic multiplets (\ref{n-arctic-1}) of weight $+2$ and $0$,
respectively, and $H^{++}$ a real $O(2)$ multiplet.

It is also instructive to express the action 
in a 4D $\cN=1$ form by switching to the two-component spinor notation
\be
\cD^{\1}_{\hal}\,=\,\left(\begin{array}{c}\cD^{\1}_{\a}\\ 
\cDB^{\1\ad}\end{array}\right)\,=\,\left(\begin{array}{c}\cD^{\1}_{\a}\\ \cDB^{\ad}_{\2}
\end{array}\right)~, \qquad 
\cD^{\2}_{\hal}\,=\,\left(\begin{array}{c}\cD^{\2}_{\a}\\ 
\cDB^{\2\ad}\end{array}\right)\,=\,\left(\begin{array}{c}\cD^{\2}_{\a}\\- \cDB^{\ad}_{\1}
\end{array}\right)~.
\ee
Using the analyticity conditions (\ref{ancon22})
we can express $\cDB^\ad_{\2}$ via $ \cDB^\ad_{\1}$.
As a result
our action 
(\ref{1projectiveActionN=1}) 
becomes
\bea
S&=&\oint {\rd\z \over 2\pi\ri\z}
\int\rd^5 x\,e\,
\bigg({1\over 16}({\cD}^{\1})^2({\cDB}_{\1})^2
-\z\o J^{\1}_{~\1}({\cD}^{\1})^2\bigg)\cL(z,\z)\Big|~.
\label{form1}
\eea
Using the identities
\bea
{[}\cDB^\ad_{\1},({\cD}^{\1})^2{]}&=&8\o J^{\1}{}_{\1}\cDB^\ad_{\2}
-4\ri(\tilde{\s}^a)^{\ad\a}\cD_a\cD^{\1}_\a
+8\o J^{\1}{}_{\1}\cD^{\1\a}M_\a{}^\ad~,\\
{[}({\cD}^{\1})^2,({\cDB}_{\1})^2{]}&=&
-\,16\o J^{\1}{}_{\1} \cDB_{\1\ad}\cDB_{\2}^\ad
+16\o J^{\1}{}_{\1} \cD^{\1\a}\cD^{\2}_\a
+96\o^2J^{\1}{}_{\1}J\non\\
&&
+\,4\ri({\s}^a)_\a^{~\ad}\cD_a{[}\cD^{\1\a},\cDB_{\1\ad}{]}
+8 \o J^{\1}{}_{\1}{[}\cD^{\1\a},\cDB_{\1\ad}{]}M_\a{}^\ad~,
\eea
the action can also be rewritten in the following form
\bea
S&=&\oint {\rd\z \over 2\pi\ri\z}
\int\rd^5 x\,e\,
\bigg({1\over 16}({\cDB}_{\1})^2({\cD}^{\1})^2
+{\o\over \z} J^{\1}_{~\1}({\cDB}_{\1})^2
\bigg)\cL(z,\z)\Big|~,
\label{form2}
\eea
or in the  manifestly real form
\bea
S&=&\oint {\rd\z \over 2\pi\ri\z}
\int\rd^5 x\,e\,
\bigg({1\over 32}\Big\{({\cD}^{\1})^2,({\cDB}_{\1})^2\Big\}
-\hf\z\o J^{\1}_{~\1}({\cD}^{\1})^2
+{\o\over 2\z} J^{\1}_{~\1}({\cDB}_{\1})^2
\bigg)\cL(z,\z)\Big|~.~~~~~~
\label{4projectiveActionN=1}
\eea
As compared with the flat superspace action \cite{KuzLin}, 
the second and third terms on the right of (\ref{4projectiveActionN=1})
are  due to the non-vanishing curvature.

\subsection{Nonlinear sigma-models}

We consider a system of interacting artic weight-one multiplets 
$\U^{+ } (z,u^+) $ and their smile-conjugates
$ \widetilde{\U}^{+}$ described by the Lagrangian
\bea
\cL^{++} = {\rm i} \, K(\U^+, \widetilde{\U}^+)~,
\label{conformal-sm}
\eea
with $K(\F^I, {\bar \F}^{\bar J}) $ a real analytic function.
Since $\cL^{++} =\cL^{++} (z,u^+) $ is required to  be a weight-two projective 
superfield, the potential  $K$ has to respect the following homogeneity condition
\be
\Big( \F^I \frac{\pa}{\pa \F^I} +  {\bar \F}^{\bar I} \frac{\pa}{\pa {\bar \F}^{\bar I}} \Big)
K(\F, \bar \F) = 2\, K( \F,   \bar \F) ~.
\label{Kkahler}
\ee
${}$For $\cal{L}^{++}$ to be real, it is sufficient to require a stronger condition
\bea
 \F^I \frac{\pa}{\pa \F^I} K(\F, \bar \F) =  K( \F,   \bar \F)~.
 \label{Kkahler2}
 \eea
 Such a Lagrangian
 corresponds to the  superconformal sigma-model 
 introduced in \cite{KuzConformal5D}.
Then, representing $\U^+(z,u^+) =u^{+\1} \,\U(z,\z) $
and $\widetilde{\U}^+(z,u^+) =u^{+\2} \,\widetilde{\U}(z,\z) $, we 
can rewrite the Lagrangian in the form
\bea
\cL^{++} (z,u^+)= {\rm i} \, u^{+\1} u^{+\2} \, \cL(z,\z) ~, \qquad 
\cL =K(\U, \widetilde{\U})~.
\eea

Because of freedom  (\ref{L++})
in the choice of Lagrangian, we can generalize 
the above construction by replacing $K(\F^I, {\bar \F}^{\bar J}) $
in (\ref{conformal-sm}) with 
\bea
K'(\F^I, {\bar \F}^{\bar J}) =K(\F^I, {\bar \F}^{\bar J}) 
 +\L(\F^I)- {\bar \L}({\bar \F}^{\bar J} ) ~,
  \qquad    \F^I \frac{\pa}{\pa \F^I} \L(\F) = 2\,\L(\F) ~,
 \eea 
 with $\L(\F)$ a holomorphic homogeneous function of degree $+2$.
Then, the homogeneity condition (\ref{Kkahler2}) turns into 
\bea
 \F^I \frac{\pa}{\pa \F^I} K'(\F, \bar \F) =  K'( \F,   \bar \F)
 +\L(\F)+{\bar \L}(\bar \F ) ~.
\eea

We can also consider a system of interacting arctic weight-zero multiplets 
${\bf \U} (z,u^+) $ and their smile-conjugates
$ \widetilde{ \bf{\U}}$ described by the Lagrangian 
\bea
\cL^{++} = \frac{\rm i}{2} \, J^{++}\,
{\bf K}({\bf \U}, \widetilde{\bf \U})~,
\eea
with ${\bf K}(\F^I, {\bar \F}^{\bar J}) $ a real function 
which is not required to obey any 
homogeneity condition. Due to the gauge freedom (\ref{L}), 
the action is invariant under K\"ahler transformations of the form
\be
{\bf K}({\bf \U}, \widetilde{\bf \U})~\to ~{\bf K}({\bf \U}, \widetilde{\bf \U})
+{\bf \L}({\bf \U}) +{\bar {\bf \L}} (\widetilde{\bf \U} )~,
\ee
with $\bf \L$ a holomorphic function. Such dynamical systems
generalize the hyperk\"ahler sigma-models on cotangent bundles 
of K\"ahler manifolds \cite{GK,AN,AKL}.

\subsection{Vector multiplet and Chern-Simons couplings}
An Abelian vector mulitplet 
can be described by 
a weight-zero real projective superfield $V(z,u^+)$
which is required to be holomorphic
on ${\mathbb C}P^1 - \{N \cup S\}$.
\be
\cD^+_{\hat \a} V (z,u^+)=0~,
\qquad 
V (z,c\,u^+) = V (z,u^+)~,\qquad c \in{\mathbb C}^*~.
\ee
In the North chart, it is characterized by the  series (\ref{VM}).
It is defined to  possess the   gauge
freedom 
\be
V ~\to~ V 
+
\l 
+ \widetilde{\l} 
~, \qquad 
 \l(z,\z) =\sum_{k=0}^{+\infty} \l_k(z) \z^k~.
\label{VMgf}
\ee
with $\l (z,u^+)$ an arctic multiplet of weight 0. 
Using considerations similar to those given in subsection 
\ref{PAP},
the field strength (compare with the flat superspace expression 
\cite{KuzConformal5D})
\be
W(z)  =- \frac{1}{ 16\pi {\rm i}} \oint 
\frac{ u^+_i\,{\rm d} u^{+i}}{(u^+ u^-)^2}  \,
\Big[(\hat{\cD}^-)^2  -12\o\, J^{--}\Big] V(z,u^+)
\label{strengt4}
\ee
can be shown to be invariant under 
the projective transformations 
(\ref{projectiveGaugeVar}).
The field strength turns out  to be invariant under the gauge transformations
(\ref{VMgf}). 
In the projective  gauge  ($u^-_{\2}=0$,  $J^{\1\1}=J^{\2\2}=0$), 
the field strength takes the form 
\be
W(z)  =- \frac{1}{ 16\pi {\rm i}} \oint 
{\rm d} \z  \,
(\hat{\cD}^{\1})^2  
V(z,\z)~,
\label{strengt5}
\ee
compare with the flat superspace result  \cite{KuzLin}.

The AdS transformation law of $V$, 
\be
\d V = -\big( \x + {\rm i}\,\r J \big) V~,
\ee
can be shown to imply that $W$ transforms as 
\be
\d W = -\x \,W
\ee
under the isometry group.

The field strength can be shown to obey the Bianchi identity
\be
\cD^{(i}_{\hat \a} \cD_{\hat \b }^{j)}  W
= {1 \over 4} \ve_{\hat \a \hat \b} \,
\cD^{\hat \g (i} \cD_{\hat \g }^{j)}  W~, 
\label{Bianchi1}
\ee
and therefore
\be
\cD^{(i}_{\hat \a} \cD_{\hat \b }^{j}  \cD_{\hat \g }^{k)} \cW
= 2\o\, \ve_{\hat \b \hat \g} \,J^{(ij} \cD^{k)}_{\hat \a} W~,
\label{Bianchi1.5}
\ee
compare with the flat superspace case \cite{Z,KuzLin}.
The Bianchi identity implies that 
\be
G^{++} (z,u^+) = 
{\rm i}\, \Big\{ \cD^{+ \hat \a} W \, \cD^+_{\hat \a} W 
+\hf  W \, 
({\hat \cD}^+)^2 W  -2\o\, J^{++} W^2 \Big\}
\label{YML}
\ee
is a composite $O(2)$ multiplet, 
\be
\cD^+_{\hat \a} G^{++}=0 ~, \qquad 
G^{++} (z,u^+) = G^{ij}(z) \,u^+_iu^+_j~.
\ee

Let $H^{++} (z,u^+) $ be a real $O(2)$ multiplet. 
Then, similarly to the flat superspace case \cite{KuzLin,KuzConformal5D}, 
the supersymmetric action associated with the Lagrangian 
\be
\cL^{++} = V(z,u^+)\,H^{++}(z,u^+)
\ee
can be shown to be invariant under the gauge transformations 
(\ref{VMgf}). 

Given several Abelian vector multiplets $V_I(z,u^+)$, 
where $I=1,\dots, n$,  the composite superfield (\ref{YML}) 
is generalised to the form: 
\bea
G^{++}_{IJ} =G^{++}_{(IJ)}
&=&{\rm i}\,
\Big\{ \cD^{+ \hat \a} W_{I} \, \cD^+_{\hat \a} W_{J} 
+\hf \,W_{(I} \,  ({\hat \cD}^+)^2 W_{J)}
-2\o\, J^{++} W_I W_J
 \Big\}~,  \non \\
\cD^+_{\hat \a} G^{++}_{IJ}&=&0 ~,  \qquad 
G^{++}_{IJ} (z,u^+) = G^{ij}_{IJ}(z) \,u^+_iu^+_j~.
\eea
We then can construct 
a supersymmetric Chern-Simons action associated with the Lagrangian  
\be
\cL^{++}_{\rm CS} =
\frac{1}{12 }\,c_{I ,JK}\,
V_I (z,u^+) \, 
G^{++}_{JK} (z,u^+) ~, 
\qquad
c_{I ,JK} =c_{I, KJ}~,
\label{CS3}
\ee
for some constant parameters $c_{I ,JK} $ 
(compare with the flat superspace case \cite{KuzLin,KuzConformal5D}). 
In accordance with the above result, the Chern-Simons  action is gauge invariant.

\subsection{Tensor multiplet  and vector-tensor couplings} 
Given several $O(2)$ 
(or, equivalently, tensor) multiplets $H^{++}_I(z,u^+)$, 
a supersymmetric action is generated by the Lagrangian
\be
\cL^{++} =\cF( H^{++}_I ) ~, \qquad I=1,\dots ,n~
\label{tm}
\ee
where  $\cF (H) $ is a weakly homogeneous function 
of first degree in the variables $H$,
\be
H_I \, \frac{\pa \cF(H  ) }{\pa H_I } 
-\cF (H  )  = \a^I \,H_I~,
\label{tm2}
\ee
for some constants $\a$'s.\footnote{The projective action principle 
formulated in subsection 5.2 requires the Lagrangian to be a projective
weight-two multiplet.  With $\a^I \neq 0$ in (\ref{tm2}), 
the Lagrangian (\ref{tm}) does not have any definite weight, 
and hence the results of subsection 5.2 are not applicable directly. 
We plan to discuss the case with $\a^I \neq 0$ 
in more detail somewhere else.}
Such a Lagrangian occurs in the models for superconformal 
 tensor multiplets in four 
\cite{dWRV} and five dimensions \cite{KuzConformal5D}. 

One can also consider systems of coupled vector and tensor multiplets
described by a Lagrangian  of the form 
\be
\cL^{++} =\cF( H^{++}_I )  +V_I \Big( \k_I  H^{++}_I 
+ \frac{1}{12 }\,c_{I ,JK}\,
G^{++}_{JK} \Big)~,
\ee
for some coupling constants $\k_I$ and $c_{I ,JK}$. 

\sect{Coset space realization}
\label{CosetSection}

In this section we would like to give an explicit realization 
for the $\cN=1$ AdS$_5$ supergeometry which we have studied  
in section \ref{sectCovariantDerivatives} 
using  the  representation-independent approach.
${}$From the group-theoretical point of view, 
it is known that  the $\cN=1$ AdS$_5$ superspace 
(or simply AdS$^{5|8}$) 
can be identified with the coset space
SU(2,2$|$1)/SO(4,1)$\times$U(1). 
Using the formalism of nonlinear realizations\footnote{Many years ago, 
this formalism was also applied to introduce the 4D $\cN=1$ 
AdS superspace \cite{4D-AdS,IS}.} 
\cite{nlr}
(or Cartan's coset construction),
here we introduce  a suitable  coset representative that makes possible
to realize one half of AdS${}^{5|8}$
as  a trivial fiber bundle with fibers isomorophic to four-dimensional Minskowski
superspace. This realization should  be useful 
if one is  interested in having the 4D $\cN=1$ super Poincar\'e symmetry manifest. 
However, since it corresponds to one half of AdS$^{5|8}$
(known as the Poincar\'e patch \cite{AGMOO}), 
it is not suitable to describe the supersymmetric actions.

The analysis of this section builds on the construction 
given in \cite{KuzMcA2001c}, see also 
\cite{BelIvaKri2002b} for related issues. 
Note that we use the superform convenctions of \cite{WessBagger}.

\subsection{Coset representative}
\label{subCoset1}

As is well known, 
the supergroup SU(2,2$|$1) is the four-dimensional
$\cN=1 $ superconformal group. It
is generated by Lie-algebra elements
of the form (parametrization (\ref{algebraSU(2,2|1)})
was used in 
\cite{OsbornConformal,KuzThe1999})
\be
X = \left(
\begin{array}{ccc}
w_\a{}^\b - \D \d_\a{}^\b  \quad &  -{\ri}  
b_{\a \bd} \quad &
2 \r_\a \\
 -{\ri} 
a^{\ad \b} \quad & -{\bar w}^\ad{}_\bd
+ {\bar \D}  \d^\ad{}_\bd   \quad &
2 {\bar \e}^{\ad}  \\
2 \e^\b \quad & 2 
{\bar \r}_{ \bd} \quad & 2({\bar \D} - \D)
\end{array}
\right)~,
\label{algebraSU(2,2|1)}
\ee
which satisfy the conditions
\be
{\rm str} \;X = 0~, \qquad
B X^\dag B = - X~, \qquad
B = \left(
\begin{array}{ccc}
0 \quad & 
\mathbbm{1}
\quad & 0 \\
\mathbbm{1} 
\quad & 0 \quad & 0 \\
0 \quad & 0 \quad & -1
\end{array}
\right)~.
\ee
The matrix elements in (\ref{algebraSU(2,2|1)}) correspond to a
4D Lorentz transformation $(w_\a{}^\b,~{\bar w}^\ad{}_\bd)$,
a translation $a^{\ad \a}$, a special conformal transformation
$ b_{\a \ad}$, a $Q$--supersymmetry $(\e^\a,~ {\bar \e}^{\ad })$,
an $S$--supersymmetry $(\r_\a,~{\bar \r}_{ \ad})$,
and a combined scale and U(1)--chiral transformation
$\D = \hf \l  +\frac{\rm i}{3} \t   $.

The explicit parametrization for the algebra su$(2,2|1)$, 
which is  given in (\ref{algebraSU(2,2|1)}),
is ideally suited to describe the compactified Minkowski space 
SU(2,2$|1) / (\cP \times {\mathbb C}^*)$, where $\cP$ denotes 
the $\cN=1$ super Poincar\'e group (generated by the parameters 
$(w_,~{\bar w}, ~ b,~ \r,~ {\bar \r} )$ in (\ref{algebraSU(2,2|1)})),
and ${\mathbb C}^*$ denotes the group of scale and chiral transformations 
generated by the parameters $\D$ and $\bar \D$ in 
(\ref{algebraSU(2,2|1)}). 
In the case of the coset space
SU(2,2$|$1)/SO(4,1)$\times$U(1), however,
this parametrization should be slightly modified.
In addition, a re-scaling of some matrix elements is needed in
order to incorporate the AdS curvature $\o^2$ into the formalism. 

As is known, a key role in the coset construction for $\cM= G/H$ 
is played by a coset representative $S(p)$ defined 
to be a smooth mapping $S$:  $U \to G$, for some open domain 
$U \subset \cM$, such that $S(p) p_0 = p,$ for any point $p\in U$, 
where $p_0 \in U$ is a fixed point having $H$ as its isotropy group. 
On topological grounds, it is not always possible to extend $U$ to 
the whole coset space $\cM$.

As a coset representative, $S(z)$, for 
AdS$^{5|8} =$ SU(2,2$|$1)/SO(4,1)$\times$U(1),
following mainly \cite{KuzMcA2001c}
we  choose 
\bea
S (z) &=&g(\bm{z})\cdot g_S\cdot g_D
\non \\
&=&
\left(
\begin{array}{ccc}
\mathbbm{1}
~ &      0  ~  &      0  \\
-{\rm i} {\o} \tilde{x}_+  ~& 
\mathbbm{1}
~ & 2 {\o}^\hf{\bar \q} \\
2 {\o}^\hf\q  ~ & 0 ~ & 1
\end{array} \right)
\left(
\begin{array}{ccc}
\mathbbm{1} ~  &    2  {\o}\eta {\bar \eta}   ~  &      2 {\o}^\hf \eta  \\
0 ~ & 
\mathbbm{1}
~ & 0 \\
0  ~ & 2 {\o}^\hf {\bar \eta}  ~& 1
\end{array} \right)
\left(
\begin{array}{ccc}
\re ^{-\hf{\o} y} 
\mathbbm{1}
~  &   0   ~  &    0 \\
0 ~ & \re^{\hf{\o} y} 
\mathbbm{1}
~ & 0 \\
0  ~ & 0   ~& 1
\end{array} \right)
\label{para}   \\
&=& \left(
\begin{array}{ccc}
\re ^{-\hf{\o} y} \d_\a^{~\b} \quad &
2 {\o}\re ^{\hf {\o}y} \eta_\a {\bar \eta}_\bd
& 2 {\o}^\hf \eta_\a \\
 -{\rm i} {\o}\re ^{-\hf {\o}y} \tilde{x}_+^{\ad\b} \quad &
\re ^{\hf {\o}y} \big(\d^\ad_{~\bd}- 2{\rm i} {\o}^2 \tilde{x}^{\ad\g}_+  
\eta_\g {\bar \eta}_\bd+4  {\o}{\bar \q}^\ad {\bar \eta}_\bd \big)
\quad &
2 {\o}^\hf\big( {\bar \q}^\ad  - {\rm i} {\o} \tilde{x}^{\ad\g}_+ \eta_\g \big) \\
2 {\o}^\hf\re ^{-\hf{\o} y}\q^\b \quad & 
2 {\o}^\hf\re ^{\hf{\o} y} \big( \bar \eta_\bd+ 2  {\o}\q^\g \eta_\g {\bar \eta}_\bd  \big) \quad &
\big( 1 + 4  {\o}\q^\g \eta_\g \big)
\end{array}
\right) ~,\non
\eea
where $x^a_\pm = x^a \pm {\rm i} \q \s^a {\bar \q}$ denote ordinary 
4D $\cN=1$ (anti) chiral bosonic variables. 
It is worth pointing out that 
the coset representative $g(\bm{z})$ 
corresponds to the coset $\cP/{\rm SO}(3,1)$ and provides a matrix 
realization\footnote{It is a curious historic fact that the above
matirx realization   for 4D $\cN=1$ Minkowski superspace
 was introduced by Akulov and Volkov 
\cite{AV}
a year before the official discovery  of superspace.}
for 4D $\cN=1$ Minkowski superspace, with coordinates
$\bm{z} = (x^a, \q^\a ,{\bar \q}_\ad )$.
Note that the isotropy group at $z=0$ is 
$H= {\rm SO}(4,1) \times {\rm U}(1) \in {\rm SU}(2,2|1)=G$, and it  is generated by
matrices of the form
\bea
\bH  = \left(
\begin{array}{ccc}
{\bf w}
~ &  -{\ri\over 2} {\bf b}
~ & 0 \\
{\ri\over 2} \tilde{{\bf b}}  ~ & - {\bf {\bar w}}
~ & 0  \\
0  ~ & 0 ~ & 0
\end{array}
\right)
+ \left(\begin{array}{ccc}
-{\ri\over 3} {\bf \tau}\,
\mathbbm{1}
~ &  0~ & 0 \\
0 ~ & -{\ri\over 3} {\bf \t}\,
\mathbbm{1}
~ & 0  \\
0  ~ & 0 ~ & -{4\ri\over 3} {\bf \t} 
\end{array}
\right)~,\qquad
\begin{array}{cccc}
&{\rm tr} \,{\bf w} =0~,&&~{\bf {\bar w}}= {\bf w}^\dag~,\vspace{0.3cm}\\ 
&{\bf b}^\dag= {\bf b}~,&&\bar{\bf\t}={\bf\t}~.
 \end{array}
\eea
Setting $\o=1$ in (\ref{para}) gives the parametrization used in \cite{KuzMcA2001c}.

Once the coset representative $S(p)$ is chosen, the next step 
in the coset construction for  $\cM= G/H$ is to compute
the Maurer-Cartan one-form $S^{-1}\rd S$ 
which proves to  encode all  the information about the geometry of $\cM$.
Let $\cG$ and $\cH$ be the Lie algebras of $G$ and $H$, respectively, 
and $\cG-\cH$ be a complement of $\cH$ in $\cG$ such that 
$[\cG-\cH, \cH ] \subset \cG-\cH$.
Then, the Maurer-Cartan one-form can be uniquely decomposed as 
 $S^{-1}\rd S =  S^{-1}\rd S |_{\cG-\cH}+ S^{-1}\rd S |_{\cH}$,
 where $S^{-1}\rd S |_{\cG-\cH}$ is identified with the vielbein, and
 $S^{-1}\rd S |_{\cH}$ with the connection.

In our case, the vielbein ${\bf E}=S^{-1}\rd S|_{\cG -\cH}$
and the connection ${\bf \O}=S^{-1}\rd S|_{\cH}$ 
are:
\bea
 S^{-1}\, {\rm d} S &=& \bE + {\bf \O}~,\non \\
\bE&=&\left(
\begin{array}{ccc}
-{1\over 2}{\o}{\bf E}_y\,\d_\a{}^\b ~ &
-{\rm i\over 2} {\o}  {\bf E} _{\a\bd} ~
& 2  {\o}^\hf({\bf E} _\eta)_\a \\
 -{\rm i\over 2} {\o} \tilde{\bf E}^{\ad\b} ~ &
{1\over 2}{\o} {\bf E}_y \,\d^{\ad}{}_\bd
~ & 2 {\o}^\hf  (\bar{\bf E}_{\bar \q})^\ad  \\
2 {\o}^\hf ({\bf E}_\q)^\b ~ & 2  {\o}^\hf(\bar{\bf E}_{\bar \eta})_\bd
~ & 0
\end{array}
\right)~,\\
{\bf \O}&=&
\left(
\begin{array}{ccc}
 {\bf \O}_\a{}^\b -{\ri\over 3}\, {\bf \O}_{\rm U(1)}\d_\a{}^\b~ &  -\frac{\rm i}{2}  {\bf \O}_{\a\bd} ~ & 0 \\
 \frac{\rm i}{2} \tilde{\bf \O}^{\ad\b} ~ & - {\bar {\bf \O}}^{\ad}{}_\bd 
 -{\ri\over 3}\, {\bf \O}_{\rm U(1)}\d^\ad{}_\bd~ & 0\\
0 ~ & 0 ~ &- {4\ri\over 3} \,{\bf \O}_{\rm U(1)}
\end{array}
\right)~.~~~~~~~
\eea
The components of the vielbein are given by the one-forms 
\begin{subequations}
\bea
{\bf E}_{\a\ad} &=&e_{\a\ad}\,\re^{- {\o}y}(1-\re ^{2 {\o}y} {\o}^2\eta^2\bar\eta^2)
+2\ri\re^{ {\o}y}\rd \eta_\a\bar\eta_\ad
+2\ri\re^{ {\o}y}\rd \bar\eta_{\ad}\eta_\a\non\\
&&+\,4\ri {\o}\re^{ {\o}y}\rd \q_\a\eta^2\bar\eta_\ad
+4\ri {\o}\re^{ {\o}y}\rd \bar\q_\ad\eta_\a\bar\eta^2~,\\
{\bf E}_y &=& {\rm d}y  +  {\rm d} \q^\mu (-2\eta_\mu)
+  {\rm d}\bar{\q}_{\dot\mu}(-2\bar{\eta}^{\dot\mu} ) ~,\\
({\bf E}_\q)^{\a} & = &{\rm d} \q^\mu\,\d_\mu^\a\re^{-\hf {\o} y} 
+ e^m(\ri {\o}\re^{-\hf {\o} y}{\bar \eta}_{\bd}\tilde{\s}_m^{\bd\a}) ~,\\
({\bf E}_\eta)^\a  &=&{\rm d} \eta^\mu\,\d_\mu^\a\re^{\hf {\o} y}
+{\rm d} \q^\mu\,\d_\mu^\a(2 {\o}\re^{\hf  {\o}y} \eta^2)
+{\rm d} {\bar \q }_{\dot\mu}(2 {\o}\re^{\hf  {\o}y}{\bar \eta}^{\dot\mu}\eta^\a ) 
+e^m(\ri  {\o}^2\re^{\hf {\o} y}\eta^2\bar\eta_\bd\tilde{\s}_m^{\bd\a})~.~~~ ~~~~~
\eea
\end{subequations}
The components of the SO(4,1)$\times$U(1) connection  read
\begin{subequations}
\bea
{\bf \O}_\a{}^\b &=& 
\rd \q^\mu (4 {\o}\eta_\a\d^\b_\mu-2 {\o}\eta_\mu\d^\b_\a)
+e^m(-2\ri {\o}^2\eta_\a\bar\eta_\bd\tilde{\s}_m^{\bd\b}
-\ri {\o}^2\bar\eta_{\dot{\g}}\tilde{\s}_m^{\dot{\g}\g}\eta_\g\d_\a^\b)~, \\
{\bf \O}_{\a\ad}  &=&-e_{\a\ad}\, {\o}\re^{- {\o}y}(1+ {\o}^2\re^{2 {\o}y}\eta^2\bar\eta^2)
+\rd \eta_\a(2\ri {\o}\re^{ {\o}y}\bar\eta_{\dot\a})
+\rd \bar\eta_{\dot\a}(2\ri {\o}\re^{ {\o}y}\eta_{\a})\non\\
&&+\,\rd \q_\a(4\ri {\o}^2\re^{ {\o}y}\eta^2\bar\eta_{\dot\a})
+\rd \bar\q_{\dot\a}(4\ri {\o}^2\re^{ {\o}y}\eta_{\a}\bar\eta^2)~,\\
{\bf \O}_{\rm U(1)}&=&\rd\q^\mu(\,3\ri {\o}\eta_\mu) 
+\rd\bar\q_{\dot\mu}(-3\ri {\o}\bar\eta^{\dot\mu})
+e^m(-3 {\o}^2\bar\eta_{\dot{\mu}}\tilde{\s}_m^{\dot{\mu}\mu}\eta_{\mu})~,
\eea
\end{subequations}
where
\bea
e^m&=&\rd x^m -\ri\,\rd\q^\mu\s^a_{\mu\dot{\mu}}\bar\q^{\dot{\mu}} 
+\ri\,\q^\mu\s^a_{\mu\dot{\mu}}\rd\bar\q^{\dot\mu}~,
\eea
is the space-time component of the  $\cN=1$ 
flat superspace vielbein \cite{WessBagger}.

Note that under a group transformation $g \in {\rm SU}(2,2|1)$
\bea
g\,S(z)\,=\,S(g\cdot z)\,\hat{h}(z;g)\,\equiv\,S'\,\hat{h}~,~~~~~~
\hat{h}(z;g)\in{H}~,
\label{groupTransform}
\eea
the vielbein ${\bf E}$ and the connection ${\bf \O}$ transform as
follows:
\bea
{\bf E}'\,=\,
\hat{h}\,{\bf E}\,\hat{h}^{-1}~,
~~~~~~{\bf \O}'\,=\,
\hat{h}\,{\bf \O}\,\hat{h}^{-1}-(\rd \hat{h})\,\hat{h}^{-1}~.
\label{connectionTransform}
\eea

It is useful to introduce the inverse $E_A{}^M$ of the vielbein supermatrix 
$E_M{}^A$ implicitly used in the previous equations
($E_A{}^ME_M{}^B\,=\,\d_A{}^B,~E_M{}^AE_A{}^N\,=\,\d_M{}^N$).
With the definitions
\bea
\varepsilon^M&=&\big(e^m,\rd y,\rd\q^\mu,
\rd\bar\q_{\dot\mu},\rd\eta^\mu,\rd\bar\eta_{\dot\mu}\big)\,
=\,{\bf E}^AE_A{}^M~,\\
{\bf E}^A&=&\big({\bf E}^a,{\bf E}_y,({\bf E}_\q)^\a,(\bar{\bf E}_{\bar\q})_\ad,
({\bf E}_\eta)^\a,(\bar{\bf E}_{\bar\eta})_\ad\big)\,=\, \varepsilon^M E_M{}^A~,
\eea
where $e^m=-\hf( \tilde{\s}^m)^{\ad\a}e_{\a\ad}$
and  $\bE^a=-\hf( \tilde{\s}^a)^{\ad\a}\bE_{\a\ad}$,
we find
\begin{subequations}
\bea
e_{\a\ad}& = & {\bf E}_{\a\ad}\,\re^{ {\o}y}(1+ {\o}^2\re^{2 {\o}y}\eta^2\bar\eta^2)
+({\bf E}_\eta)_\a(-2\ri\re^{{3\over 2}  {\o}y}\bar\eta_\ad)
+\,(\bar{\bf E}_{\bar\eta})_\ad(-2\ri\re^{{3\over 2} {\o} y}\eta_\a)\non\\
&&+\,({\bf E}_\q)_\a(2\ri {\o}\re^{{5\over 2}  {\o}y}\bar\eta_\ad\eta^2)
+\,(\bar{\bf E}_{\bar\q})_\ad(2\ri {\o}\re^{{5\over 2}  {\o}y}\eta_\a\bar\eta^2)~,\\
\rd y & = & {\bf E}_y
+({\bf E}_\q)^\a(\,2\re^{\hf {\o} y}\eta_\a)
+(\bar{\bf E}_{\bar\q})_\ad(\,2\re^{\hf {\o} y}\bar\eta^\ad)~,\\
\rd \q^{\mu} & = & ({\bf E}_\q)^\a\,\re^{\hf {\o} y}\d_\a^\mu(1-2 {\o}^2\re^{2 {\o}y}\eta^2\bar\eta^2)
+({\bf E}_\eta)^\a\,\d^\mu_\a(2 {\o}\re^{{3\over 2}  {\o}y}\bar\eta^2)\non\\
&&+\,(\bar{\bf E}_{\bar\eta})_\ad(-2 {\o}\re^{{3\over 2} {\o} y}\bar\eta^\ad\eta^\mu)
+{\bf E}^a(-\ri {\o}\re^{ {\o}y}\bar\eta_{\dot{\nu}}\tilde{\s}_a^{\dot{\nu}\mu})~,\\
\rd \eta^{\mu} & = & ({\bf E}_\eta)^\a\,\re^{-\hf  {\o}y}\d_\a^\mu(1-4 {\o}^2\re^{2 {\o}y}\eta^2\bar\eta^2)
+({\bf E}_\q)^\a\,\d_\a^\mu(-2 {\o}\re^{\hf {\o} y}\eta^2)\non\\
&&+\,(\bar{\bf E}_{\bar\q})_\ad(-2 {\o}\re^{\hf  {\o}y}\bar\eta^\ad\eta^\mu)
+{\bf E}^a(2\ri {\o}^2\re^{ {\o}y}\bar\eta_{\dot{\nu}}\tilde{\s}_a^{\dot{\nu}\mu}\eta^2)~.
\eea
\end{subequations}

It is also useful to decompose the connection 
with respect to the curved basis  $\{ {\bf E}^A \}$ 
\begin{subequations}
\bea
{\bf \O}_\a^{~\b}&=&({\bf E}_\q)^\g\, {\o}\re^{\hf {\o} y}(4\eta_\a\d^\b_\g-2\eta_\g\d_\a^\b)
+({\bf E}_\eta)^\g\, {\o}\re^{{3\over 2} {\o} y}(4\eta_\a\bar\eta^2\d^\b_\g
-2\eta_\g\bar\eta^2\d_\a^\b)\non\\
&&+\,{\bf E}^a\, {\o}^2\re^{ {\o}y}(-2\ri\bar\eta_{\bd}\tilde{\s}_a^{\bd\b}\eta_\a
+\ri\bar\eta_{\dot{\g}}\tilde{\s}_a^{\dot{\g}\g}\eta_\g\d^\b_\a)
~,\\
{\bf \O}_{\a\ad}&=&{\bf E}_{\a\ad}\, {\o}(-1-2 {\o}^2\re^{2 {\o}y}\eta^2\bar\eta^2)
+({\bf E}_\eta)_\a(4\ri {\o}\re^{\hf {\o} y}\bar\eta_\ad)
+\,(\bar{\bf E}_{\bar\eta})_\ad(4\ri {\o}\re^{\hf  {\o}y}\eta_\a)\non\\
&&+\,({\bf E}_\q)_\a(-4\ri {\o}^2\re^{{3\over 2}  {\o}y}\bar\eta_\ad\eta^2)
+\,(\bar{\bf E}_{\bar\q})_\ad(-4\ri {\o}^2\re^{{3\over 2}  {\o}y}\eta_\a\bar\eta^2)~,\\
{\bf \O}_{\rm U(1)}&=&{\bf E}^a(\,3 {\o}^2\re^{ {\o}y}\bar\eta_\bd\tilde{\s}_a^{\bd\b}\eta_\b)
+({\bf E}_\q)^\a(\,3\ri {\o}\re^{\hf  {\o}y}\eta_\a)
+(\bar{\bf E}_{\bar\q})_{\dot\a}(-3\ri {\o}\re^{\hf  {\o}y}\bar\eta^{\dot\a})\non\\
&&+\,({\bf E}_\eta)^\a(\,6\ri {\o}^2\re^{{3\over 2}  {\o}y}\eta_\a{\bar\eta}^2)
+(\bar{\bf E}_{\bar\eta})_{\dot\a}(-6\ri {\o}^2\re^{{3\over 2} {\o}y}\bar\eta^{\dot\a}\eta^2)~.
\eea
\end{subequations}

\subsection{SO(4,1)$\times$U(1) covariance}
\label{subsectCovariance}

To better understand the relation between the above coset construction 
and the AdS$^{5|8}$ supergeometry of section
\ref{sectCovariantDerivatives}, it is necessary 
to figure out the precise meaning of the SO(4,1)$\times$U(1) 
covariance of the vielbein 
and the connection.  We will use several results 
which are collected in  Appendix \ref{sect5DConvenctions} 
and concern  the reduction of 5D spinors into 4D ones.

${}$First of all, let us recall that choosing 
$g=h \in H$ in relations (\ref{groupTransform}, \ref{connectionTransform}) 
gives ${\hat h} = h ={\rm const}$,
and the group transformations 
(\ref{connectionTransform}) reduce to 
\bea
{\bf E}'\,=\,h\,{\bf E}\,h^{-1}~,~~~{\bf \O}'\,=\,h\,{\bf \O}\,h^{-1}~,
~~~~~~h\in{\rm SO(4,1)}\times{\rm U(1)}~.
\label{IsometryTransformation}
\eea
In particular, a 5D Lorentz transformation acts as follows:
\bea
{\bf E}'\,=\,\L\,{\bf E}\,\L^{-1}~,~~~{\bf \O}'\,=\,\L\,{\bf \O}\,\L^{-1}~,\label{5DLorentz}
\eea
where
\be
\L\,=\,\left(\begin{array}{c|c}
\L_{\hal}{}^{\hbe}~&~
\begin{matrix}0\,\\0\,\end{matrix}\\\hline
\begin{matrix}0&&0\end{matrix}&\,1
\end{array}
\right)~,~~~~~~
\L_{\hal}{}^{\hbe}\,=\,
\Big{[}\exp\Big( \hf\L^{{\hat c} {\hat d}}(\S_{{\hat c}{\hat d}})\Big)\Big{]}_{\hal}{}^{\hbe}~.
\ee
This transformation law allows us to combine  components of the connection 
into five-dimensional  vector and spinor.  Explicitly, we can write 
\be
{\bf E}\,=\,\left(\begin{array}{c|c}
-{\ri\over 2} {\o}\,{\bf E}^{\ha}(\G_{\ha})_{\hal}{}^{\hbe}~&~
2 {\o}^\hf{\bf E}_{\hal}\begin{matrix}\,\\\,\end{matrix}\\\hline
2 {\o}^\hf\bar{\bf E}^{\hbe}&0
\end{array}
\right)~,\label{5Dvielbein}
\ee
\be
{\bf \O}\,=\,{1\over 2} \,{\bf \O}^{\ha\hb}\left(\begin{array}{c|c}
(\S_{\ha\hb})_{\hal}{}^{\hbe}~&~
\begin{matrix}0\,\\0\,\end{matrix}\\\hline
\begin{matrix}0~~&~~0\end{matrix}&\,0
\end{array}
\right)~+~
\ri\,{\bf \O}_{\rm U(1)}\left(\begin{array}{c|c}
-{1\over 3}\d_{\hal}{}^{\hbe}~&~
\begin{matrix}0\,\\0\,\end{matrix}\\\hline
\begin{matrix}0&&0\end{matrix}&\,-{4\over 3}
\end{array}
\right)~,\label{5Dconnection}
\ee
where
\begin{subequations}
\bea
&{\bf E}^{\ha}\,=\,({\bf E}^{a},{\bf E}^{5})\,=\,({\bf E}^a,{\bf E}_y)~,\label{E5D}\\
&{\bf E}_{\hal}\,=\,\left(\begin{array}{c}({\bf E}_\eta)_\a\\ 
(\bar{\bf E}_{\bar{\q}})^\ad\end{array}\right)~,~~~
\bar{{\bf E}}^{\hal}=\,\Big(({\bf E}_\q)^\a,(\bar{\bf E}_{\bar\eta})_\ad\Big)~,\\
&{\bf \O}^{\ha\hb}\,=\,({\bf \O}^{ab},{\bf \O}^{a5})~,\\
&{\bf \O}^{ab}\,=\,-(\s^{ab})_\b{}^\a{\bf \O}_\a{}^\b
+(\tilde\s^{ab})^\bd{}_\ad{}\bar{\bf \O}^\ad{}_\bd~,~~~
{\bf \O}^{a5}\,=\,-\hf(\tilde{\s}^a)^{\ad\a}{\bf \O}_{\a\ad}~.
\eea
\end{subequations}
Note that ${\bf E}^{\ha}$, ${\bf \O}_{\ha\hb}=-{\bf \O}_{\hb\ha}$ and ${\bf \O}_{\rm U(1)}$ are real.
It follows that ${\bf E}^{\ha}$, ${\bf E}_{\hal}$, $\bar{\bf E}^{\hal}$, 
${\bf \O}_{\ha\hb}$ and ${\bf \O}_{\rm U(1)}$ transform 
under the 5D Lorentz group ${\rm SO}(4,1)$
respectively as a vector, 
a Dirac spinor, its Dirac conjugate spinor, an antisymmetric two-tensor and a scalar.
Due to  (\ref{E5D}) we can identify
\be
x^5~\equiv~y~.
\ee

Note also that we can combine  the two spinors ${\bf E}_{\hal}$ 
and $\bar{\bf E}^{\hal}$ into a 5D pseudo-Majorana spinor defined as follows:
\bea
&{\bf E}^{\hal}_i\,=\,({\bf E}^{\a}_i,-\bar{\bf E}_{i\ad})~,\\
&{\bf E}^{\a}_{\1}\,=\,({\bf E}_\q)^\a~,~~~
{\bf E}^{\a}_{\2}\,=\,({\bf E}_\eta)^\a~,~~~
\bar{\bf E}^{\1}_{\ad}\,=\,(\bar{\bf E}_{\bar{\q}})_\ad~,~~~
\bar{\bf E}^{\2}_{\ad}\,=\,(\bar{\bf E}_{\bar{\eta}})_\ad~.
\eea

It remains to  consider the transformation properties of the vielbein 
and the connection under the U(1) part of the isotropy group.
In accordance with (\ref{IsometryTransformation}), they transform as
\bea
&{\bf E}'\,=\,
\S\,{\bf E}\,
\S^{-1}~,~~~
{\bf \O}'\,=\,
\S\,{\bf \O}\,
\S^{-1}~, \non \\
&
\S\,=\,
\left(\begin{array}{c|c}
\left[\exp(-{1\over 3}\f\,\ri\,\d)\right]_{\hal}{}^{\hbe}&~
\begin{matrix}0\,\\0\,\end{matrix}\\\hline
\begin{matrix}0&~~~&0\end{matrix}&\,\re^{-{4\over 3}\f\,\ri}
\end{array}
\right)~.\label{U(1)matrix}
\eea
Clearly ${\bf \O}$ is invariant under the U(1) transformation,
while ${\bf E}$  transforms as
\be
{\bf E}'\,=\,\left(\begin{array}{c|c}
-{\ri\over 2} {\o}\,{\bf E}^{\ha}(\G_{\ha})_{\hal}{}^{\hbe}~&~
2 {\o}^\hf\big({\rm e}^{\f\ri}\,{\bf E}_{\hal}\big)\begin{matrix}\,\\\,\end{matrix}\\\hline
2 {\o}^\hf\big({\rm e}^{-\f\ri}\,\bar{\bf E}^{\hbe}\big)&0
\end{array}
\right)~,
\ee
and hence ${\bf E}^{\ha}$ is invariant. Note also that 
(\ref{U(1)matrix}) induces
induces the following transformation of  ${\bf E}^i_{\hal}$:
\be
{\bf E}'{}_i^{\hal}\,=\,\left[\exp(-\f\ri J)\right]_i{}^{j}{\bf E}_j^{\hal}~,~~~~~~
J_i{}^{j}\,=\,(\s_3)_i{}^{j}\,=\,\left(\begin{matrix}1~&\,0\\ 0~&-1\end{matrix}\right)~.
\label{Jcoset}
\ee

\subsection{Representation of covariant derivatives}

With the vielbein and the connection having been introduced, 
we can now construct the 
covariant derivatives 
\bea
\cD_{\hat{A}}&=&E_{\hat{A}}+{\rm i}\,\Phi_{\hat{A}}\,J+\hf \,\O_{\hat{A}}{}^{\hb\hc}\,M_{\hb\hc}
\,=\,E_{\hat{A}}+{\rm i}\,\Phi_{\hat{A}}\,J+\hf \,\O_{\hat{A}}{}^{bc}\,M_{bc}
+\O_{\hat{A}}{}^{b5}\,M_{b5}\non\\
&=&(\cD_{\ha},\cD^i_{\hal})\,=\,(\cD_a,\,\cD_5,\,\cD^{\1}_\a,\,
 \bar{\cD}^{\1\ad},\,\cD^{\2}_\a,\, \bar{\cD}^{\2\ad})~.
\eea
The vector fields
$E_{\hat{A}}$ are defined by
\bea
E_{\hat{A}}&=&\big(E_{\ha}, E_{\hal}^i\big)\,=\,
E_{\hat{A}}{}^MD_M~,
\non \\
D_M&=&\Big( \pa_m,{\pa\over\pa y}, D_\mu,\bar{D}^{\dot{\mu}},{\pa\over\pa\eta^\mu},{\pa\over\pa\bar \eta_{\dot{\mu}}}\Big)\,=\,E_M{}^{\hat{A}}{E}_{\hat{A}}~.
\eea
Here the supermatrices $E_{\hat{A}}{}^M$ and $E_M{}^{\hat{A}}$ 
have been defined in subsection \ref{subCoset1}.
It should be pointed out  that 
$(\pa_m,D_\mu,\bar{D}^{\dot{\mu}})$
are the 4D $\cN=1$ flat superspace covariant derivatives,
$D_\mu={\pa\over\pa\q^\mu}+\ri\bar\q^{\dot{\mu}}\pa_{\mu\dot{\mu}}$ and 
$\bar{D}^{\dot{\mu}}={\pa\over\pa\bar\q_{\dot{\mu}}}+\ri\q_{\mu}\tilde{\pa}^{\dot{\mu}\mu}$. 
Furthermore, the connection supefields in $\cD_{\hat{A}}$ are defined as
\bea
{\bf\O}_{\rm U(1)}\,=\,\bE^{\hat{A}}\,\F_{\hat{A}}~,
~~~~~~{\bf \O}^{\ha\hb}\,=\,\bE^{\hat{A}}\,\O_{\hat{A}}{}^{\ha\hb}~.
\eea 
It can be shown that the explicit expressions for the covariant derivatives are as follows:
\begin{subequations}
\bea
\cD_a&=&\re^{{\o}y}(1+{\o}^2\re^{2{\o}y}\eta^2\bar\eta^2)\pa_a
-\ri{\o}\re^{{\o}y}(\bar\eta\tilde{\s}_a)^\mu D_\mu
-\ri{\o}\re^{{\o}y}(\eta\s_a)_{\dot\mu} \bar D^{\dot\mu}
\non\\
&&
+\,2\ri{\o}^2\re^{{\o}y}\eta^2(\bar\eta\tilde{\s}_a)^{\mu}{\pa\over\pa\eta^\mu}
+\,2\ri{\o}^2\re^{{\o}y}\bar\eta^2(\eta{\s}_a)_{\dot\mu}{\pa\over\pa\bar\eta_{\dot\mu}}\non\\
&&-\,3\ri{\o}^2\re^{{\o}y}(\eta{\s}_a\bar\eta)J
+{\o}^2\re^{{\o}y}\eta_{ab}\ve^{bcde}(\eta\s_c\bar\eta)M_{de}
-{\o}(1+2{\o}^2\re^{2{\o}y}\eta^2\bar{\eta}^2)M_{a5}~,\label{covDev1}\\
\cD_5&=&{\pa\over\pa y}~,\label{covDev2}\\
\cD^{\1}_\a&=&\re^{\hf{\o} y}(1-2{\o}^2\re^{2{\o}y}\eta^2\bar\eta^2)D_\a
-2{\o}\re^{\hf{\o} y}\eta^2{\pa\over\pa\eta^\a}
-2{\o}\re^{\hf {\o}y}\eta_\a\bar\eta_{\dot\mu}{\pa\over\pa\bar\eta_{\dot\mu}}\non\\
&&+\,2\eta_\a\re^{\hf {\o}y}{\pa\over \pa y}
-\ri{\o}\re^{{5\over 2}{\o} y}\eta^2(\s^m\bar\eta)_\a\pa_m\non\\
&&-\,3{\o}\re^{\hf {\o}y}\eta_\a\,J+
2{\o}\re^{\hf {\o}y}\eta^\b(\s^{ab})_{\b\a}M_{ab}+
2\ri{\o}^2\re^{{3\over 2}{\o}y}\eta^2(\s^a\bar{\eta})_\a M_{a5}~,
\label{covDev3}\\
\cD^{\2}_\a&=&\re^{-\hf {\o}y}(1-4{\o}^2\re^{2{\o}y}\eta^2\bar\eta^2){\pa\over\pa\eta^\a}
+2{\o}\re^{{3\over 2}{\o} y}\bar\eta^2D_\a
-2{\o}\re^{{3\over 2} {\o}y}\eta_\a\bar\eta_{\dot\mu}\bar D^{\dot\mu}
+\ri\re^{{3\over 2} {\o}y}(\s^m\bar\eta)_\a\pa_m\non\\
&&-\,6{\o}^2\re^{{3\over 2} {\o}y}\eta_\a{\bar\eta}^2J
+2{\o}^2\re^{{3\over 2} {\o}y}\eta^\b\bar{\eta}^2(\s^{ab})_{\b\a}M_{ab}
-2\ri{\o}\re^{{1\over 2}{\o}y}(\s^a\bar{\eta})_\a M_{a5}~,
\label{covDev4}\\
\cDB_{\1}^\ad&=&\re^{\hf {\o}y}(1-2{\o}^2\re^{2{\o}y}\eta^2\bar\eta^2)\DB^\ad
-2{\o}\re^{\hf {\o}y}\bar{\eta}^2{\pa\over\pa\bar\eta_\ad}
-2{\o}\re^{\hf {\o}y}\bar{\eta}^\ad\eta^{\mu}{\pa\over\pa\eta^{\mu}}\non\\
&&+\,2\bar{\eta}^\ad\re^{\hf {\o}y}{\pa\over \pa y}
-\ri{\o}\re^{{5\over 2}{\o} y}\bar{\eta}^2(\tilde{\s}^m\eta)^\ad\pa_m\non\\
&&+\,3{\o}\re^{\hf{\o} y}\bar{\eta}^\ad\,J
+2{\o}\re^{\hf{\o} y}\bar{\eta}_\bd(\tilde{\s}^{ab})^{\bd\ad}M_{ab}
+2\ri{\o}^2\re^{{3\over 2}{\o}y}\bar{\eta}^2(\tilde{\s}^a{\eta})^\ad M_{a5}~,
\label{covDev3bar}\\
\cDB_{\2}^\ad&=&\re^{-\hf {\o}y}(1-4{\o}^2\re^{2{\o}y}\eta^2\bar\eta^2){\pa\over\pa\bar\eta_\ad}
+2{\o}\re^{{3\over 2} {\o}y}\eta^2\DB^\ad
-2{\o}\re^{{3\over 2}{\o} y}\bar\eta^\ad\eta^{\mu}D_{\mu}
+\ri\re^{{3\over 2} y}(\tilde{\s}^m\eta)^\ad\pa_m\non\\
&&+\,6{\o}^2\re^{{3\over 2} {\o}y}\bar\eta^\ad{\eta}^2\,J
+2{\o}^2\re^{{3\over 2}{\o} y}\bar\eta_\bd{\eta}^2(\tilde{\s}^{ab})^{\bd\ad}M_{ab}
-2\ri{\o}\re^{{1\over 2}{\o}y}(\tilde{\s}^a{\eta})^\ad M_{a5}~.
\label{covDev4bar}
\eea
\end{subequations}

It is interesting to consider a flat superspace limit, $\o \to 0$, for the covariant derivatives.
In this limit,  one finds
\bea
\cD_{\hat{A}}\Big|_{\o \to 0} &=&
{\rm e}^{-U}  D_{\hat{A}} \,{\rm e}^U~, \qquad
U = \eta \q +{\bar \eta} \bar \q~, 
\eea
where $D_{\hat A} = ( \pa_{\hat a} , D^i_{\hat \a}) $ are 5D flat global 
covariant derivatives, 
\be
D^i_{\hat \a} = \frac{\pa}{\pa \q^{\hat \a}_i} 
- {\rm i} \, (\G^{\hat b} ){}_{\hat \a \hat \b} \, \q^{\hat \b i}
\, \pa_{\hat b}~,
\ee
with $\q^{\hat \a}_i = ( \q^\a_i , - {\bar \q}_{ \ad i})$ and  
$\q^\a_i =(\q^\a, \eta^\a)$.

\subsection{Torsion and curvature}

Now, we are prepared to demonstrate that the geometry 
described in the present section 
reproduces  the geometry of ${\rm AdS}^{5|8}$
constructed in section \ref{sectCovariantDerivatives}.

We proceed by recalling that, in accordance with the  coset construction, 
the torsion ${\bf T}$ and curvature ${\bf R}$ two-forms are defined as follows:
\bea
{\bf T}\,=\,
\rd {\bf E}\,-\,{\bf \O}\wedge{\bf E}\,-\,{\bf E}\wedge{\bf \O}~,~~~~~~
{\bf R}\,=\,
\rd {\bf \O}\,-\,{\bf \O}\wedge{\bf \O}~.
\eea
Under group transformations (\ref{groupTransform}) they transform covariantly 
\bea
{\bf T}'\,=\,
\hat{h}\,{\bf T}\,{\hat h}^{-1}~,~~~~~~
{\bf R}'\,=\,
\hat{h}\,{\bf R}\,{\hat h}^{-1}~.
\label{curvatureTransform}
\eea
Keeping in mind the definition ${\bf  E}+{\bf \O}= G^{-1}\rd G$, we get
\bea
\rd{\bf  E}+\rd{\bf \O}
\,=\,G^{-1}\rd G\wedge G^{-1}\rd G
&=&{\bf  E}\wedge{\bf  E}\,+\,{\bf  E}\wedge{\bf  \O}\,
+\,{\bf  \O}\wedge{\bf  E}\,+\,{\bf  \O}\wedge{\bf  \O}~,
\eea
from which we obtain
\bea
\rd  {\bf  E}\,=\,
({\bf  E} \wedge {\bf  E})|_{\cG -\cH}\,+\,{\bf  E}\wedge{\bf  \O}\,+\,{\bf  \O}\wedge{\bf  E}~,~~~~~~
\rd {\bf  \O}\,=\,
({\bf  E}\wedge{\bf  E})|_{\cH}\,+\,{\bf  \O}\wedge{\bf  \O}~,
\eea
since $({\bf  E}\wedge{\bf  \O}+{\bf  \O}\wedge{\bf  E})\in{\cG -\cH}$ 
and ${\bf  \O}\wedge{\bf  \O}\in{\cH}$.
Using the previous formulae
 we are able to see that the torsion and 
curvature two-forms 
are given by simple expressions
\bea
{\bf T}\,=\,
({\bf  E}\wedge{\bf  E})|_{\cG-\cH}~,~~~~~~
{\bf R}\,=\,
({\bf  E}\wedge{\bf  E})|_{\cH}~.
\label{TorsionCurvature}
\eea
Therefore, it remains to compute ${\bf  E}\wedge{\bf  E}$.

Direct calculations give
\be
{\bf E}\wedge\bE\,=\,\left(\begin{array}{c|c}
\hf\,{\o}^2\,{\bf E}^{\ha}\wedge\bE^{\hb}(\S_{\ha\hb})_{\hal}{}^{\hbe}
+4{\o}\,\bE_{\2\hal}\wedge\bE_{\1}^{\hbe}~&
-\ri{\o}^{3\over 2}\,\bE^{\ha}\wedge{\bf E}_{\2\hga}\,(\G_{\ha})_{\hal}{}^{\hga}\begin{matrix}\,
\\\,\end{matrix}\\\hline
-\ri{\o}^{3\over 2}\,{\bf E}_{\1}^{\hga}\wedge\bE^{\hb}\,
(\G_{\hb})_{\hga}{}^{\hbe}&4{\o}\,{\bf E}_{\1}^{\hga}\wedge\bE_{\2\hga}
\end{array}
\right)~,
\ee
and this we should represent as 
${\bf E}\wedge\bE =({\bf  E}\wedge{\bf  E})|_{\cG -\cH}+
({\bf  E}\wedge{\bf  E})|_{\cH}$.
We end up with 
\be
{\bf T}\,=\,\left(\begin{array}{c|c}
-{\ri\over 2}{\o}\,{\bf T}^{\ha}(\G_{\ha})_{\hal}{}^{\hbe}~&~2{\o}^\hf\,{\bf T}_{\2\hal}\begin{matrix}\,\\\,\end{matrix}\\\hline
2{\o}^\hf\,{\bf T}_{\1}^{\hbe}&0
\end{array}
\right)~,\label{torsion2}
\ee
\be
{\bf R}\,=\,{1\over 2}\,\bR^{\ha\hb}\left(\begin{array}{c|c}
(\S_{\ha\hb})_{\hal}{}^{\hbe}~&~
\begin{matrix}0\,\\0\,\end{matrix}\\\hline
\begin{matrix}0&~~~~&0\end{matrix}&\,0
\end{array}
\right)~+~
\ri\,\bR_{\rm U(1)}\left(\begin{array}{c|c}
-{1\over 3}\d_{\hal}{}^{\hbe}~&~
\begin{matrix}0\,\\0\,\end{matrix}\\\hline
\begin{matrix}0&&0\end{matrix}&\,-{4\over 3}
\end{array}
\right)~,\label{curvature2}
\ee
where
\bea
\bT^{\ha}&=&{1\over 2}\,\bE_{k}^{\hga}\wedge\bE_{j}^{\hbe}\,\Big(2\ri\,\ve^{jk}
(\G^{\ha})_{\hbe\hga}\Big)~,\label{torsion2.1}\\
\bT_{i}^{\hal}&=&\hf\Bigg{[}\,\bE^{\hc}\wedge{\bf E}_{j}^{\hbe}\,
\Big(\,{\ri\over 2}{\o}(\s_3)^{~j}_{i}(\G_{\hc})_{\hbe}{}^{\hal}\Big)
+{\bf E}_{k}^{\hga}\wedge\bE^{\hb}\,\Big(-{\ri\over 2}{\o}(\s_3)^{~k}_{i}
(\G_{\hb})_{\hga}{}^{\hal}\Big)\Bigg{]}~,~~~
\label{torsion2.2}
\\\non\\
\bR^{\ha\hb}&=&\hf\Bigg{[}\,{\bf E}^{\hat{d}}\wedge\bE^{\hat{c}}\,{\o}^2
(-\d^{\ha}_{\hc}\d^{\hb}_{\hat{d}}+\d^{\hb}_{\hc}\d^{\ha}_{\hat{d}})
+\,\bE_{l}^{\hde}\wedge\bE_{k}^{\hga}\,\Big(-4{\o}\ve^{ki}(\s_3)^{~l}_{i}
(\S^{\ha\hb})_{\hga\hde}\Big)\Bigg{]}~,\label{curvature2.1}\\
\bR_{\rm U(1)}&=&\hf\,\bE_{l}^{\hde}\wedge{\bf E}_{k}^{\hga}\,
\Big(3\ri{\o}\ve^{kl}\ve_{\hga\hde}\Big)~\label{curvature2.2}.
\eea
Using standard superform definitions \cite{WessBagger},
we  define the components  of the torsion and curvature 
as follows:
\bea
&\bT^{{\hat{A}}}\,=\,
\hf\,\bE^{{\hat{C}}}\wedge\bE^{\hat{B}}\, T_{\hat{B}\hat{C}}{}^{\hat{A}}~,\\
&\bR^{\ha\hb}\,=\,\hf\,\bE^{{\hat{D}}}\wedge\bE^{\hat{C}}\,R_{\hat{C}\hat{D}}{}^{\ha\hb}~,~~~
\bR_{\rm U(1)}\,=\,\hf\,\bE^{{\hat{D}}}\wedge\bE^{\hat{C}}\,(R_{\rm U(1)})_{\hat{C}\hat{D}}~.
\eea

Now, let us return 
to the covariant derivatives described in section 
\ref{sectCovariantDerivatives}.  
Their algebra given by eqs. (\ref{algebra1}--\ref{algebra3})  can 
be represented concisely  as
\bea
\big{[}\cD_{\hat{A}},\cD_{\hat{B}}\big{\}}\,=\,-T_{\hat{A}\hat{B}}{}^{\hat{C}}\cD_{\hat{C}}
+\,\ri\,(R_{\rm U(1)})_{\hat{A}\hat{B}}\,J
+\hf\,R_{\hat{A}\hat{B}}{}^{\hc\hat{d}}M_{\hc\hat{d}}~.
\eea
Comparing (\ref{algebra1}--\ref{algebra3})  
with eqs. (\ref{torsion2.1}--\ref{curvature2.2}), we find that 
all the components  of the torsion and curvature coincide provided
\bea
J^i_{~j}\,=\,(\s_3)^i_{~j}~.
\eea
This completes our analysis of the coset construction.
\\

\noindent
{\bf Acknowledgements:}\\
This work is supported  in part
by the Australian Research Council and by a UWA research grant.

\appendix

\sect{5D Conventions}
\label{sect5DConvenctions}

Our 5D notation and conventions correspond to \cite{KuzLin}.
The 5D gamma-matrices $\G_{\hat m} = ( \G_m, \G_5 )$, 
with $m=0,1,2,3$,
are defined by 
\be
\{ \G_{\hat m} \, , \,\G_{\hat n} \} 
= - 2 \eta_{\hat m \hat n} \,
\mathbbm{1}
~, \qquad
(\G_{\hat m} )^\dagger = \G_0 \, \G_{\hat m} \, \G_0 
\ee
are chosen in accordance with 
\bea
(\G_m ){}_{\hat \a}{}^{\hat \b}=
\left(
\begin{array}{cc}
0 ~ &  (\s_m)_{\a\bd} \\
(\tilde{\s}_m)^{\ad \b} ~ & 0  
\end{array}
\right)~, \qquad
(\G_5 ){}_{\hat \a}{}^{\hat \b}=
\left(
\begin{array}{cc}
-{\rm i} \,\d_\a{}^\b~ &  0 \\
0 ~ & {\rm i}\, \d^{ \ad}{}_{\bd}
\end{array}
\right)~,\label{Gamma1}
\eea
such that $\G_0 \G_1 \G_2 \G_3 \G_5 =
\mathbbm{1}$. 
The charge conjugation matrix, $C = (\ve^{\hat \a \hat \b})$, 
and its inverse, $C^{-1} = C^\dag =(\ve_{\hat \a \hat \b})$ 
are defined by 
\bea
C\,\G_{\hat m} \,C^{-1} = (\G_{\hat m}){}^{\rm T}~,
\qquad 
\ve^{\hat \a \hat \b}=
\left(
\begin{array}{cc}
 \ve^{\a \b} &0 \\
0& -\ve_{\ad \bd}    
\end{array}
\right)~, \quad
\ve_{\hat \a \hat \b}=
\left(
\begin{array}{cc}
 \ve_{\a \b} &0 \\
0& -\ve^{\ad \bd}    
\end{array}
\right)~.
\eea
The antisymmetric matrices $\ve^{\hat \a \hat \b}$ and
$\ve_{\hat \a \hat \b}$ are used to raise and lower the four-component 
spinor indices.

A Dirac spinor, $\J=(\J_{\hat \a}) $, and its Dirac conjugate, 
$\bar \J =({\bar \J}^{\hat \a}) = \J^\dag \,\G_0$, look like
\bea
\J_{\hat \a} =   
\left(
\begin{array}{c}
\j_\a \\
{\bar \f}^{\ad}    
\end{array}
\right)~, \qquad 
{\bar \J}^{\hat \a}= (\f^\a \,, \,{\bar \j}_{\ad})~.
\eea
One can now combine ${\bar \J}^{\hat \a}= (\f^\a , {\bar \j}_{\ad})$ and 
$\J^{\hat \a} = \ve^{\hat \a \hat \b} \J_{\hat \b} =(\j^\a , - {\bar \f}_{\ad} )$ 
into a SU(2) doublet, 
\be
\J^{\hat \a}_i = (\J^\a_i,  -{\bar \J}_{\ad i} ) ~, \qquad
(\J^\a_i)^* = {\bar \J}^{\ad i}~, \qquad 
i = \1 , \2 ~,   
\ee 
with $\J^\a_{\1} = \f^\a $ and $\J^\a_{\2} = \j^\a $.
It is understood that the SU(2) indices are raised and lowered 
by $\ve^{ij} $ and  $\ve_{ij} $, $\ve^{\1 \2} =  \ve_{\2 \1} =1$, 
in the standard fashion: $\J^{\hat \a i} = \ve^{ij} \J^{\hat \a}_j$.
The  Dirac  spinor $\J^i = ( \J^i_{\hat \a}  )$
satisfies the pseudo-Majorana condition
${\bar \J}_i{}^{\rm T} = C \,   \J_i$.
This will be concisely represented as
\be
(\J^i_{\hat \a} )^* = \J^{\hat \a}_i~.
\ee

With the definition $\S_{\hat m \hat n} 
=-\S_{\hat n \hat m} = -{1 \over 4}
[\G_{\hat m} , \G_{\hat n} ] $, the matrices 
$\{ \mathbbm{1}, \G_{\hat m} , \S_{\hat m \hat n} \} $
form a basis in the space of  $4 \times 4$ matrices. 
The matrices $\ve_{\hat \a \hat \b}$ and 
$(\G_{\hat m})_{\hat \a \hat \b}$ are antisymmetric, 
$\ve^{\hat \a \hat \b}\, (\G_{\hat m})_{\hat \a \hat \b} =0$, 
while the matrices $(\S_{\hat m \hat n})_{\hat \a \hat \b}$ 
are symmetric.  

It is useful to write explicitly the 4D reduction of these matrices
\bea
(\G_m )_{\hat \a\hat \b}\,=&
\left(
\begin{array}{cc}
0 ~ &-  (\s_m)_\a{}^\bd \\
({\s}_m)_\b{}^\ad ~ & 0  
\end{array}
\right)~, ~~~
(\G_5 ){}_{\hat \a\hat \b}&=\,
\left(
\begin{array}{cc}
{\rm i} \,\ve_{\a\b}~ &  0 \\
0 ~ & {\rm i}\, \ve^{ \ad\bd}
\end{array}
\right)~,\label{Gamma2}\\
\non\\
(\S_{mn})_{\hat \a}{}^{\hat \b}\,=&
\left(
\begin{array}{cc}
(\s_{mn})_\a{}^\b~ &0 \\
 0~ & (\tilde{\s}_{mn})^\ad{}_\bd  
\end{array}
\right)~, ~~~
(\S_{m5} )_{\hat \a}{}^{\hat \b}&=\,
\left(
\begin{array}{cc}
0~ &  -{\ri\over 2}(\s_m)_{\a\bd} \\
 {\ri\over 2}(\tilde{\s}_m)^{\ad\b}~ &0
\end{array}
\right)~,~~~\label{Gamma3}\\
\non\\
(\S_{mn})_{\hat \a\hat \b}\,=&
\left(
\begin{array}{cc}
(\s_{mn})_{\a\b}~ &0 \\
 0~ & -(\tilde{\s}_{mn})^{\ad\bd}  
\end{array}
\right)~,~~~
(\S_{m5} )_{\hat \a\hat \b}&=\,
\left(
\begin{array}{cc}
0~ &  {\ri\over 2}(\s_m)_\a{}^\bd \\
 {\ri\over 2}({\s}_m)_\b{}^\ad~ &0
\end{array}
\right)~,~~~\label{Gamma4}
\eea
where $(\s_{mn})_\a{}^\b=-{1\over 4}(\s_m\tilde{\s}_n-\s_n\tilde{\s}_m)_\a{}^\b$ and
$(\tilde{\s}_{mn})^\ad{}_\bd=-{1\over 4}(\tilde{\s}_m\s_n-\tilde{\s}_n\s_m)^\ad{}_\bd$.

Given a 5-vector $V^{\hat m}$ and an 
antisymmetric tensor $F^{\hat m \hat n} = -F^{\hat n \hat m}$,
we can equivalently represent  them as the 
bi-spinors $V = V^{\hat m} \,  \G_{\hat m}$
and $F = \hf F^{\hat m \hat n}\, \S_{\hat m \hat n} $
with the following symmetry properties
\bea 
V_{\hat \a \hat \b} &=& -V_{\hat \b \hat \a} ~, 
\quad \ve^{\hat \a \hat \b}\, V_{\hat \a \hat \b} =0~, 
 \qquad \quad
F_{\hat \a \hat \b} = F_{\hat \b \hat \a} ~. 
\eea
The two equivalent descriptions 
$ V_{\hat m} \leftrightarrow V_{\hat \a \hat \b}$ and 
and $ F_{\hat m \hat n} \leftrightarrow F_{\hat \a \hat \b}$
are explicitly described as follows:
\bea 
V_{\hat \a \hat \b} = V^{\hat m} \, ( \G_{\hat m})_{\hat \a \hat \b}~,
\quad && \quad 
V_{\hat m}  = -{1 \over 4} \,( \G_{\hat m})^{\hat \a \hat \b}\,
V_{\hat \a \hat \b}~, \non \\
F_{\hat \a \hat \b} = \hf F^{\hat m \hat n} 
(\S_{\hat m \hat n})_{\hat \a \hat \b}~, \quad && \quad 
F_{\hat m \hat n}  = (\S_{\hat m \hat n})^{\hat \a \hat \b} \,
F_{\hat \a \hat \b} ~.
\eea
These results can be easily checked using the identities
\bea
 \ve _{\hat \a \hat \b \hat \g \hat \d} 
&=& \ve_{\hat \a \hat \b} \, \ve_{\hat \g \hat \d}
+ \ve_{\hat \a \hat \g} \, \ve_{\hat \d \hat \b}
+\ve_{\hat \a \hat \d} \, \ve_{\hat \b \hat \g}~,  \non \\
 \ve_{\hat \a \hat \g} \, \ve_{\hat \b \hat \d}
-\ve_{\hat \a \hat \d} \, \ve_{\hat \b \hat \g}
&=&-\hf \,  ( \G^{\hat m})_{\hat \a \hat \b}\,
( \G_{\hat m})_{\hat \g \hat \d}
+\hf \, \ve _{\hat \a \hat \b}\, \ve_{ \hat \g \hat \d} ~,
\eea
and therefore 
\be
 \ve _{\hat \a \hat \b \hat \g \hat \d} 
=\hf \,( \G^{\hat m})_{\hat \a \hat \b}\,
( \G_{\hat m})_{\hat \g \hat \d}
+\hf \, \ve _{\hat \a \hat \b} \, \ve_{ \hat \g \hat \d} ~,
\ee
with 
$ \ve _{\hat \a \hat \b \hat \g \hat \d} $ the completely 
antisymmetric fourth-rank tensor.

Complex conjugation gives 
\be 
(\ve_{\hat \a \hat \b})^* = - \ve^{\hat \a \hat \b}~,
\qquad 
(V_{\hat \a \hat \b})^* = V^{\hat \a \hat \b}~,
\qquad 
(F_{\hat \a \hat \b})^* = F^{\hat \a \hat \b}~,
\ee
provided   $V^{\hat m}$ and  $F^{\hat m \hat n} $ are real.

\small{

}

\end{document}